\documentstyle[12pt]{article}
\newcommand{\nsection}{\setcounter{equation}{0}
\def\theequation{\thesection.\arabic{equation}}\section}

\newcommand{\appendixA}{\setcounter{equation}{0}
\def\theequation{\rm{A}.\arabic{equation}}\section*}
\newcommand{\appendixB}{\setcounter{equation}{0}
\def\theequation{\rm{B}.\arabic{equation}}\section*}

\def\simlt{\stackrel{<}{{}_\sim}}
\def\simgt{\stackrel{>}{{}_\sim}}

\newcount\hour
\newcount\minute
\newtoks\amorpm
\hour=\time\divide\hour by60
\minute=\time{\multiply\hour by60 \global\advance\minute by-\hour}
\edef\standardtime{{\ifnum\hour<12 \global\amorpm={am}%
        \else\global\amorpm={pm}\advance\hour by-12 \fi
        \ifnum\hour=0 \hour=12 \fi
        \number\hour:\ifnum\minute<10 0\fi\number\minute\the\amorpm}}
\edef\militarytime{\number\hour:\ifnum\minute<10 0\fi\number\minute}

\def\be{\begin{equation}}
\def\ee{\end{equation}}
\def\bear{\be\begin{array}}
\def\eear{\end{array}\ee}
\def\bea{\begin{eqnarray}}
\def\eea{\end{eqnarray}}

\def\sb{\sin\beta}
\def\cb{\cos\beta}

\def\c2bb{\cos^2 2 \beta}

\def\s2bb{\sin^2 2 \beta}

\def\simlt{\stackrel{<}{{}_\sim}}
\def\simgt{\stackrel{>}{{}_\sim}}
\def\ov{\overline}

\def\baselinestretch{1}
\begin{document}

\begin{titlepage}

\title{{\bf The lightest Higgs boson mass in the Minimal
Supersymmetric Standard Model}\thanks{Work partly supported
by CICYT under contract AEN94-0928, and  by the European Union under contract
No. CHRX-CT92-0004.}
}

\author{{\bf J.A. Casas${}^{1,2}$},
{\bf J. R. Espinosa${}^{2}$}\thanks{
Supported by a grant of Comunidad de Madrid, Spain.},
{\bf M. Quir\'os${}^{2}$}
and {\bf A. Riotto${}^{2}$}\thanks{On leave of absence
from International School of Advanced Studies, ISAS, Trieste.
}\\
\hspace{3cm}\\
${}^{1}$CERN, TH Division, CH-1211 Geneva 23, Switzerland\\
\hspace{3cm}\\
${}^{2}$Instituto de Estructura de la Materia, CSIC\\
Serrano 123, 28006-Madrid, Spain}

\date{}
\maketitle
\def\baselinestretch{1.15}
\begin{abstract}
\noindent
We compute the upper bound on the mass of the lightest Higgs boson
in the Minimal Supersymmetric Standard Model in a model-independent way,
including leading (one-loop)
and next-to-leading order (two-loop) radiative corrections. We find
that (contrary to some recent claims)
the two-loop corrections are negative
with respect to the one-loop result and relatively small
($\simlt 3$\%). After defining physical (pole) top quark mass
$M_t$, by including QCD self-energies, and physical Higgs
mass $M_H$, by including the electroweak self-energies
$\Pi\left(M_H^2\right)-\Pi(0)$, we obtain the upper limit
on $M_H$ as a function of supersymmetric parameters. We include as
supersymmetric parameters the scale of supersymmetry
breaking $M_S$, the value of $\tan \beta$ and
the mixing between stops $X_t= A_t + \mu \cot\beta$
(which is responsible for the threshold correction on the
Higgs quartic coupling). Our results do not depend on further details
of the supersymmetric model. In particular, for
$M_S\leq 1$ TeV, maximal threshold effect $X_t^2=6M_S^2$
and any value of $\tan\beta$,
we find $M_H\leq 140$ GeV for $M_t\leq 190$ GeV. In the particular
scenario where the top is in its infrared fixed point
we find $M_H\leq 86$ GeV for $M_t = 170$ GeV.
\end{abstract}

\thispagestyle{empty}

\leftline{}
\leftline{CERN--TH.7334/94}
\leftline{July 1994}
\leftline{}

\vskip-22.cm
\rightline{}
\rightline{{\bf CERN--TH.7334/94}}
\rightline{{\bf IEM--FT--87/94}}
\rightline{{\bf hep-ph/9407389}}
\vskip3in

\end{titlepage}
\newpage
\setcounter{page}{1}

\nsection{Introduction}

There are good reasons to believe that the
Standard Model (SM) is not the ultimate theory
since it is unable to answer many fundamental questions.
One of them, why and how the electroweak
and the Planck scales are so hierarchically separated, has motivated
the proposal of the Minimal Supersymmetric extension of the Standard Model
(MSSM) as the underlying theory at scales of order 1 TeV. Indeed, the
supersymmetric scale cannot be very large not to spoil the solution to
the hierarchy problem, but the ambiguity about its maximal allowable size
makes it very difficult to give model--independent predictions testable
in accelerators.

Fortunately, the supersymmetric predictions for the upper bound on the
lightest Higgs boson mass represent a
(perhaps unique) exception to this rule.
Therefore, they are crucial for the experimental verification of
supersymmetry. This importance is reinforced by the fact that
uncovering the Higgs boson of the SM is one of the main
challenges for present (LEP, Tevatron) and future (LEP-200, LHC)
accelerators.
Much work has been recently devoted to this subject [1--10], but still
there is a substantial disagreement on the final results (see {\em e.g.}
[5,9,10]), especially concerning the two--loop corrections (which, as
we will see, are crucial).

The aim of this paper is to evaluate the two-loop predictions of
the MSSM on the lightest Higgs mass in a
consistent and model-independent way.
In fact, our results will not depend on the details of the MSSM ({\em
e.g.} the assumption or not of universality of the soft breaking terms,
gauge and Yukawa unification, etc.). Actually, when we will refer
to the MSSM, we will simply mean the supersymmetric version of
the SM with minimal particle content. The consistency of the approach
will allow us to show up the reasons of the disagreement between
previous results.
We also give the predictions on the Higgs mass for a particularly
appealing scenario, namely the assumption that the recently
detected top quark~\cite{cdf} has the Yukawa coupling in its infrared fixed
point. It will turn out that in this case the Higgs boson should be
just around the corner.

\vspace{0.3cm}
Let us briefly review the status of the subject and the most recent
contributions on it.
The MSSM has an extended Higgs sector with two Higgs doublets with
opposite hypercharges: $H_1$, responsible for the mass of the charged
leptons and the down-type quarks, and $H_2$, which gives a mass to the
up-type quarks. After the Higgs mechanism there remain three physical scalars,
two CP-even and one CP-odd Higgs bosons. In particular, the lightest
CP-even Higgs boson mass satisfies the tree-level bound
\begin{equation}
m_H^2\leq M_Z^2\cos^2 2\beta,
\label{bound}
\end{equation}
where $\tan\beta=v_2/v_1$ is the ratio of the Vacuum Expectation Values
(VEV's) of the neutral components of the two Higgs
fields $H_2$ and $H_1$, respectively. The relation (\ref{bound})
implies that $m_H^2<M_Z^2$, for any value of $\tan\beta$ which, in turn,
implies that it should be found at LEP-200~\cite{lep200}. However, the
tree level relation (\ref{bound}) is spoiled by one-loop radiative
corrections, which were computed by several  groups using:
the effective potential approach~\cite{eff}, diagrammatic methods
\cite{diag} and renormalization group (RG) techniques~\cite{rg}. All
methods found excellent agreement with each other and large radiative
corrections, mainly controlled by the top Yukawa coupling, which
could make the lightest CP-even Higgs boson to escape experimental
detection at LEP-200. In particular, the RG approach (which will be
followed in the present paper) is based on the
fact that supersymmetry decouples and below the scale of
supersymmetry breaking $M_S$ the effective theory is the SM.
Assuming $M_Z^2\ll M_S^2$ the tree-level bound
(\ref{bound}) is saturated at the scale $M_S$ and the effective SM at
scales between $M_Z$ and $M_S$ contains the Higgs doublet
\be
\label{hsm}
H=H_1\cb +i \sigma_2 H_2^*\sb,
\ee
with a quartic coupling taking, at the scale $M_S$, the (tree level) value
\be
\label{lms}
\lambda={\displaystyle\frac{1}{4}} (g^2+g'^2)\c2bb.
\ee
In these analyses~\cite{rg} the Higgs mass was considered at
the tree-level, improved by one-loop renormalization group equations
(RGE) in the $\gamma $-- and
$\beta $--functions, thus collecting all leading logarithm corrections.

Since the relative size of one-loop corrections to the Higgs mass is
large (mainly for large top quark mass and/or small tree level Higgs
mass) it was compelling to analyze them at the two-loop level. A
first step in that direction was given in ref.~\cite{eq} where
two-loop RGE improved tree level Higgs masses were considered. It was
found that two-loop corrections were negative and small. In ref.~\cite{eq}
the Higgs mass received all leading logarithm and part of the
next-to-leading logarithm corrections. Subsequent studies of the
effective potential improved by the RGE~\cite{kaste,bando,ford}
have shown that the
L-loop improved effective potential with (L+1)-loop RGE is exact up
to Lth-to-leading logarithm order~\cite{bando}\footnote{Strictly
speaking this has been proven \cite{bando} for a theory with
a single mass scale.}. This means that for
fully taking into account all next-to-leading logarithm corrections the
one-loop effective potential (improved by two-loop RGE) is needed.

Finally, two papers~\cite{jap,hemp} have recently appeared aiming to refine
the two-loop analysis of ref.~\cite{eq}. The authors of ref.~\cite{jap},
following the RG approach, find that the two-loop correction to the Higgs
mass is positive and sizeable, whereas the authors of ref.~\cite{hemp},
following diagrammatic \cite{diag} and effective potential~\cite{eff}
approaches in the framework of the MSSM with various
approximations, find that the two-loop correction to the Higgs
mass is negative and sizeable in contradiction with ref.~\cite{jap}.
As we will see in detail in section 5 we are in disagreement with the
results of ref.~\cite{jap}
(showing how a correct treatment of all the relevant effects
would make the results presented in ref.~\cite{jap} to agree with ours),
and in agreement with the overall result of ref.~\cite{hemp} (though not
with the relative size of the two-loop corrections).

\vspace{0.3cm}
In this paper we use the RG approach and the SM one-loop improved
effective potential
with two-loop RGE to compute the lightest Higgs mass of the MSSM up to
next-to-leading logarithm order. At this level of approximation,
one has to keep control on the validity of the perturbative expansion
for $V$. In particular, although the whole effective potential is
scale-invariant, the one-loop approximation is not, thus one has to
be careful about the choice of the renormalization scale. As we will
see this fact is at the origin of some misunderstandings in previous
works. Furthermore, since the Higgs mass is computed
to one-loop order there are a number of one-loop effects that need to
be considered for the consistency of the procedure:

{\bf i)} As it is explained in Appendix A, the tree level quartic
coupling (\ref{lms}) receives one-loop threshold contributions at the
$M_S$ scale. These are given by
\be
\label{deltal}
\Delta\lambda={\displaystyle\frac{3 h_t^4}{16
\pi^2}}{\displaystyle\frac{X_t^2}{M_S^2}}\left(2-
{\displaystyle\frac{X_t^2}{6 M_S^2}}\right),
\ee
where $h_t$ is the top Yukawa coupling in the SM and
\be
\label{Xt}
X_t=A_t+\mu\cot\beta
\ee
is the stop mixing.

The correction (\ref{deltal}) has a maximum for $X_t^2=6M_S^2$.
For that reason, in our numerical applications we will take
two cases: $X_t=0$, {\em i.e.} no mixing, and $X_t^2=6M_S^2$, {\em
i.e.} maximal threshold effect. Notice also that $X_t^2=6M_S^2$
is barely consistent with the bound from color conserving
minimum ~\cite{color}, so the case of maximal threshold really represent
a particularly extreme situation.

In addition to the previous effect, there appear effective
higher order operators ($D\geq 6$), which for $M_S\geq 1$ TeV
turn out to be negligible (for details see Appendix A).

{\bf ii)} One-loop contributions to the top quark self-energy relating the
running top mass $m_t$ to the (physical) propagator pole top mass
$M_t$. We will find this effect gives a negative and sizeable
contribution to the Higgs mass

{\bf iii)} One-loop contributions to the Higgs self-energy relating also
the running Higgs mass $m_H$ to the physical Higgs mass $M_H$
(for details see Appendix B). We
will find scale dependent contributions to the self-energy,
removing the scale dependence of $M_H$, and other scale independent
contributions. This effect will be found to be positive in all cases,
partially cancelling the previous effect.

The contents of this paper are as follows: In section 2 we present
and analyze the one-loop effective potential, for the effective SM in
the range between $M_Z$ and $M_S$, improved by the RGE. We compare
different treatments of the effective potential and define the
running Higgs masses. Our treatment of the effective potential  is
presented in detail in section 3, and the numerical results are
presented in section 4, where we make use of effects (i)-(iii) above.
Numerical and conceptual comparison with other recent approaches is
presented in section 5, and our conclusions are drawn in section 6.
In Appendix B we present analytic expressions for the one-loop
contribution to the Higgs boson self-energy $\Pi(q^2)-\Pi(0)$,
computed in the 't Hooft-Landau gauge and using the $\ov{MS}$
renormalization scheme. We also comment on the introduction of the
Higgs sector both in the effective potential and in the Higgs boson
self-energy.

\nsection{The one-loop effective potential}

Our starting point in this section will be the effective potential
of the Standard Model. This can be written in the 't Hooft-Landau
gauge as~\cite{ford}
\be
\label{veff}
V_{eff}(\mu(t),\lambda_i(t);\phi(t))\equiv V_0 + V_1 + \cdots \;\; ,
\ee
where $\lambda_i\equiv (g,g',\lambda,h_t,m^2)$ runs over all
dimensionless and dimensionful couplings and $V_0$, $V_1$ are respectively
the tree level potential and the one-loop correction, namely
\be
\label{v0}
V_0=-{\displaystyle\frac{1}{2}}m^2(t)\phi^2(t) +
{\displaystyle\frac{1}{8}}\lambda(t)\phi^4(t),
\ee
\bear{cc}
\label{v1}
V_1&={\displaystyle\frac{1}{64\pi^2}}\left\{ 6
m^4_W(t)\left[\log{\displaystyle\frac{m^2_W(t)}{\mu^2(t)}}
-{\displaystyle\frac{5}{6}}\right]+
3m_Z^4(t)\left[\log{\displaystyle\frac{m^2_Z(t)}{\mu^2(t)}}
-{\displaystyle\frac{5}{6}}\right]\right. \vspace{.5cm}\\
&\left.-12m_t^4(t)\left[
\log{\displaystyle\frac{m^2_t(t)}{\mu^2(t)}} -{\displaystyle\frac{3}{2}}
\right]\right\},
\eear
where we have used the $\ov{MS}$ renormalization scheme
\cite{msbar}. The parameters $\lambda(t)$ and $m(t)$ are the Standard
Model quartic coupling and mass, running with the RGE, while the
running Higgs field is
\be
\label{field}
\phi(t)=\xi(t) \phi_c,
\ee
$\phi_c$ being the classical field and
\be
\label{xi}
\xi(t)=e^{-\int^t_0 \gamma(t')dt'},
\ee
where $\gamma(t)$ is the Higgs field anomalous dimension.
$V_1$ in eq.(\ref{v1}) contains the radiative corrections where
only the top quark and the $W,\ Z$ gauge bosons are propagating in
the loop. For the moment we will disregard
those coming from the Higgs and Goldstone boson
propagation\footnote{In particular, the Goldstone boson contribution
to the effective potential generates, on the running Higgs mass,
an infrared logarithmic divergence. This divergence is cancelled when
the physical Higgs mass is considered, as will be shown in Appendix B.}.
They can be easily introduced, as shown in Appendix B, without
altering the numerical results that are obtained in this paper.
The mass parameters in (\ref{v1}) are given by
\bear{cl}
\label{mass}
m_W^2(t)&={\displaystyle\frac{1}{4}}g^2(t)\phi^2(t),\vspace{.5cm}\\
m_Z^2(t)&={\displaystyle\frac{1}{4}}[g^2(t)+g'^2(t)]\phi^2(t),\vspace{.5cm}\\
m_t^2(t)&={\displaystyle\frac{1}{2}}h_t^2(t)\phi^2(t),\\
\eear
where $g,\ g'$ and $h_t$ are the $SU(2),\ U(1)$ and top Yukawa
coupling, respectively. Finally the scale $\mu(t)$ is related to the
running parameter $t$ by
\be
\label{mu}
\mu(t)=\mu e^t,
\ee
where $\mu$ is a fixed scale, that we will take equal to the physical
$Z$ mass, $M_Z$.

The complete effective potential
$V_{eff}(\mu(t),\lambda_i(t);\phi(t))$ and its $n-th$ derivative
are scale-independent (see {\em e.g.} ref.~\cite{ford}), {\em i.e.}
\be
\label{scind}
\frac{d V^{(n)}_{eff}}{d t}=0\ ,
\ee
where
\be
\label{nder}
V^{(n)}_{eff}\equiv\xi^n(t){\displaystyle\frac{\partial^n}{\partial\phi(t)^n}}
V_{eff}(\mu(t),\lambda_i(t);\phi(t)).
\ee

The above property allows in principle to fix a different scale for
each value of the classical
field $\phi_c$, {\em i.e.} $\mu(t)=f(\phi_c)$ or, equivalently,
$t=t(\phi_c)=\log(f(\phi_c)/\mu)$.
In that case $V_{eff}$ becomes
$V_{eff}(\phi_c)\equiv V_{eff}(f(\phi_c),\lambda_i(t(\phi_c))$.
Then, from (\ref{scind}) and (\ref{nder}) one can readily prove that
\be
\label{nderinv}
\left. V^{(n)}_{eff}\right|_{t=t(\phi_c)}=
{\displaystyle\frac{d^n V_{eff}(\phi_c)}{d\phi_c^n}},
\ee
as expected.
This procedure has been used in some
previous works\footnote{{\em E.g.} in studies of the effective
potential stability~\cite{ford} to control large logarithms that can arise for
large field values. It has also been used in ref.~\cite{jap},
but as we will see, it
is an unnatural complication in this context.}. However, one should notice
that even though the whole effective potential is scale invariant, the
one-loop approximation is {\em not}. Therefore one would need a criterium to
fix the function $f(\phi_c)$: the only possible one would be to
minimize the radiative corrections improving, so, perturbative expansion.
Assuming that only one field, say $f$ with squared mass $m_f^2(t)$,
is contributing to the
one-loop radiative corrections, or that the latter are dominated by
this field, the most natural
choice would be $\mu^2(t)=m_f^2(t)$. Nevertheless this approach has
(in our context) some drawbacks:

\begin{description}

\item[i)] The $t$--dependence of $m_f^2(t)$ is implicit through
the RGE and so the function $t=t(\phi_c)$ cannot be
explicitly written.

\item[ii)] In most cases, in particular in our one-loop
correction (\ref{v1}), one cannot assume, within the required degree
of accuracy, that loop corrections are dominated by only one field.

\item[iii)] When computing $V^{(n)}_{eff}(\phi_c)$ in (\ref{nderinv}) the
vacuum energy $\Omega(\lambda_i(t),\mu(t))$ has to be specified since
it also acquires a $\phi_c$--dependence through the variable change
$t=t(\phi_c)$. Furthermore, the whole dependence on $\phi_c$ becomes
now much more involved.

\item[iv)] The last, but not the least, drawback is that once one has chosen
a particular function $t(\phi_c)$, one looses track of scale invariance and
so there is no way of checking how good the approximation is at the minimum
of the effective potential.

\end{description}

\noindent For the above mentioned reasons we will keep $t$ and $\phi_c$ as
independent variables. We will minimize the effective
potential (\ref{veff}), truncated at one-loop,
at some fixed scale $t=t^*$, {\em i.e.}
\be
\label{vmin}
{\displaystyle\left.\frac{\partial V_{eff}}{\partial\phi(t^*)}
\right|_{\phi(t^*)=\langle\phi(t^*)\rangle}}=0,
\ee
which will determine the VEV $\langle\phi(t^*)\rangle$. Our criterium to
fix the scale $t^*$ will be explained in the next section.
The scale independence of the whole effective potential implies that
\be
\label{derphi}
{\displaystyle\frac{d}{dt}}{\displaystyle\frac{\langle\phi(t)\rangle}
{\xi(t)}}=0.
\ee
Therefore, assuming that $t^*$ lies in the region where the one-loop
approximation to the
effective potential is reliable, the VEV of the field at
any scale can be obtained through\footnote{Notice that
$\langle\phi(t)\rangle$ defined by (\ref{vevstar1}) coincides with the value
of $\phi(t)$ which minimizes the whole effective potential. In the
one-loop approximation that we are using this is no longer true, as
we will see later on.}
\be
\label{vevstar1}
\langle\phi(t)\rangle=\langle\phi(t^*)\rangle
{\displaystyle\frac{\xi(t)}{\xi(t^*)}}.
\ee
Accordingly, we must impose
\be
\label{vevstar}
v=\langle\phi(t_Z)\rangle=\langle\phi(t^*)\rangle
{\displaystyle\frac{\xi(t_Z)}{\xi(t^*)}},
\ee
where $t_Z$ is defined as $\mu(t_Z)=M_Z$ and
\be
\label{vev}
v=(\sqrt{2}G_{\mu})^{-1/2}=246.22\ {\rm GeV},
\ee
is the ``measured" VEV for the Higgs field\footnote{We are
neglecting here one-loop electroweak radiative corrections
to the muon $\beta$-decay slightly
modifying the relation (\ref{vev}), see {\em e.g.}
ref.~\cite{peris}. We thank S. Peris for
a discussion on this point.} ~\cite{ara}.

We can trade $\langle\phi(t^*)\rangle$ by
$m^2(t^*)$ from the condition of minimum (\ref{vmin}),
which translates, using (\ref{vevstar}), into the boundary
condition for $m^2(t)$:
\bear{cc}
\label{bcm}
{\displaystyle\frac{m^2(t^*)}{v^2}}&=
{\displaystyle\frac{1}{2}}\lambda(t^*)\xi^2(t^*)
+ {\displaystyle\frac{3}{64\pi^2}}\xi^2(t^*)\left\{
{\displaystyle\frac{1}{2}}g^4(t^*)\left[\log{
\displaystyle\frac{g^2(t^*)\xi^2(t^*)v^2}{4\mu^2(t^*)}}
-{\displaystyle\frac{1}{3}}\right]\right.\vspace{.5cm}\\
&+{\displaystyle\frac{1}{4}}\left[g^2(t^*)+g'^2(t^*)\right]^2
\left[\log{\displaystyle
\frac{\left[g^2(t^*)+g'^2(t^*)\right]\xi^2(t^*)v^2}{4\mu^2(t^*)}}
-{\displaystyle\frac{1}{3}}\right]\vspace{.5cm}\\
&-\left.4h_t^4(t^*)\left[\log{
\displaystyle\frac{h_t^2(t^*)\xi^2(t^*)v^2}{2\mu^2(t^*)}}
-1\right]\right\}.
\eear

We can define now $m_H(t^*)$ as the second derivative of the effective
potential at the minimum,
evaluated at the scale $t^*$, {\em i.e.}
\bea
\label{mhstar}
m^2_H(t^*)&=&
{\displaystyle\left.\frac{\partial^2 V_{eff}}{\partial\phi(t^*)^2}
\right|_{\phi(t^*)=\langle\phi(t^*)\rangle}}\nonumber \\
&=&\lambda(t^*)\xi^2(t^*) v^2
+ {\displaystyle\frac{3}{64\pi^2}}\xi^2(t^*)v^2\left\{
g^4(t^*)\left[\log{\displaystyle\frac{g^2(t^*)\xi^2(t^*)v^2}{4\mu^2(t^*)}}
+{\displaystyle\frac{2}{3}}\right]\right.\vspace{.5cm}\nonumber \\
&+&{\displaystyle\frac{1}{2}}\left[g^2(t^*)+g'^2(t^*)\right]^2
\left[\log{\displaystyle
\frac{\left[g^2(t^*)+g'^2(t^*)\right]\xi^2(t^*)v^2}{4\mu^2(t^*)}}
+{\displaystyle\frac{2}{3}}\right]\vspace{.5cm}\nonumber \\
&-&\left.8h_t^4(t^*)\log{\displaystyle\frac{h_t^2(t^*)\xi^2(t^*)v^2}{2\mu^2(t^*)}}
\right\},
\eea
where we have used (\ref{bcm}).
At an arbitrary scale $t$ we can use the scale independence property
of the whole potential, see (\ref{nder}), and write
\be
\label{mhrun}
m_H^2(t)=m_H^2(t^*){\displaystyle\frac{\xi^2(t^*)}{\xi^2(t)}},
\ee
which will be our definition of running mass.

In the region where our approximated effective potential is scale invariant,
definition (\ref{mhrun}) should be equivalent to taking the second derivative
with respect to $\phi(t)$, evaluate it at
$\langle\phi(t)\rangle$ and use, for the VEV,
the relation (\ref{vevstar1}), {\em i.e.}
\bea
\label{mhder}
m^2_{H,der}(t)&=&
{\displaystyle\left.\frac{\partial^2 V_{eff}}{\partial\phi(t)^2}
\right|_{\phi(t)=\langle\phi(t)\rangle}}\nonumber \\
&=&-m^2(t)+{\displaystyle\frac{3}{2}}\xi^2(t)v^2\left\{\lambda(t)+
{\displaystyle\frac{1}{8\pi^2}}\left\{{\displaystyle\frac{1}{8}}g^4(t)
\left[3\log
{\displaystyle\frac{g^2(t)\xi^2(t)v^2}{4\mu^2(t)}}+1
\right]\right.\right.\vspace{.5cm}\nonumber \\
&+&{\displaystyle\frac{1}{16}}\left[g^2(t)+g'^2(t)\right]^2
\left[3\log{\displaystyle\frac{
\left[g^2(t)+g'^2(t)\right]\xi^2(t)v^2}{4\mu^2(t)}}+1\right]
\vspace{.5cm}\nonumber \\
&-&\left.\left.h_t^4(t)\left[3\log
{\displaystyle\frac{h_t^2(t)\xi^2(t)v^2}{2\mu^2(t)}}-1
\right]\right\}\right\}.
\eea
Of course, exact scale invariance would imply that (\ref{mhrun})
and (\ref{mhder})
are equal. This will allow to cross check the reliability of our approach.

\nsection{Our approach}

In this section we will describe our approach to the problem of determination
of the Higgs mass (\ref{mhrun}). The effective potential is written in
(\ref{v0}) and (\ref{v1}), and the parameters on which it depends,
$\lambda(t),\
 g(t),\ g'(t),\ g_3(t),\ h_t(t),\ \xi(t)$
and $m^2(t)$, satisfy a system of coupled
RGE with $t$--dependence governed by $\beta_{\lambda},\ \beta_{g},\
\beta_{g'},\ \beta_{g_3},\ \beta_{h_t},\ \gamma,\ \beta_{m^2}$, {\em i.e.} the
corresponding $\beta$--functions and anomalous dimension of the Higgs field,
which are evaluated to one- or two-loop order~\cite{ford}. In fact, we will
often consider two cases:
\begin{description}
\item[a)] The {\em one-loop} case,
where $\beta$ and $\gamma$--functions are
considered to one-loop and the effective potential is approximated by the
tree level term (\ref{v0}).
In this case the leading logarithms are resummed to all-loop
in the effective potential.
The Higgs mass includes then all-loop leading
logarithm contributions \cite{kaste,bando}.
\item[b)] The {\em two-loop} case,
where $\beta$ and $\gamma$--functions in the RGE
are considered to two-loop order and the effective potential is
considered in the one-loop approximation, (\ref{v0}) and (\ref{v1}).
In this case the leading and next-to-leading logarithms are resummed
to all-loop in the effective potential.
The Higgs mass includes then all-loop leading and next-to-leading
logarithm contributions \cite{bando}.
\end{description}

We will impose as boundary conditions:
\begin{itemize}
\item For the gauge couplings:
\bear{c}
\label{bcg}
g(M_Z)=0.650,\vspace{.5cm}\\
g'(M_Z)=0.355,\vspace{.5cm}\\
g_3(M_Z)=1.23\ \ .\\
\eear
\item For the top Yukawa coupling:
\be
\label{bct}
h_t(m_t)={\displaystyle\frac{\sqrt{2}m_t}{v}}\ \ ,
\ee
where $m_t$ is the running mass: $m_t(\mu(t)=m_t)=m_t$.

\item For the quartic coupling:
\be
\label{bcl}
\lambda(M_S)={\displaystyle\frac{1}{4}}
\left[g^2(M_S)+g'^2(M_S)\right]\c2bb
+{\displaystyle\frac{3h^4_t(M_S)}{16\pi^2}} \left(2
{\displaystyle\frac{X_t^2}{M_S^2}}
-{\displaystyle\frac{X_t^4}{6 M_S^4}} \right).
\ee
\item For the mass $m^2(t)$ we will take as boundary condition the value
determined by eq.~(\ref{bcm}) at the scale $t^*$.
\end{itemize}
\noindent Note that the previous boundary conditions depend on the values of
$m_t$, $t^*$ ({\em i.e.} the minimization scale) and the supersymmetric
parameters $M_S$, $X_t$ and $\tan \beta$.

Our main task now will be to determine the optimal minimization scale
$t^*$ as a function of $m_t$. Our criterion
will be that $t^*$ be in the region where the effective
potential is more scale independent. We estimate this in the following
way. For fixed $t^*$ and $m_t$, all
boundary conditions (\ref{bcg})--(\ref{bcl}) are determined and the
effective potential (\ref{veff}) can be computed for
any value of $t$. Thus the corresponding value of $\phi(t)$,
say $\phi_{{\rm min}}(t)$, that minimizes the one-loop potential
can be numerically evaluated. Had we
considered the whole effective potential, its scale independence
property (\ref{derphi}) would imply that
$\phi_{\rm min}(t)/\xi(t)=
\langle\phi(t)\rangle/\xi(t)$,
where $\langle\phi(t)\rangle$ has been defined in (\ref{vevstar1}), is
$t$--independent. Since the one-loop approximation is not exactly
scale independent the most appropriate scale is the scale $t_s$ for
which $\phi_{\rm min}(t)/\xi(t)$ has a stationary point, {\em i.e.}
\be
\label{derphitm}
\left.
{\displaystyle\frac{d}{dt}}{\displaystyle\frac{\phi_{{\rm min}}(t)}{\xi(t)}}
\right|_{t=t_s}=0.
\ee
Of course the scale at which (\ref{derphitm}) occurs depends on
$t^*$ and $m_t$, {\em i.e.} $t_s\equiv t_s(t^*,m_t)$. Therefore, for a
given value of $m_t$ the optimal scale $t^*$ is defined as
\be
\label{tstar}
t_s\equiv t_s(t^*,m_t)=t^*,
\ee
{\em i.e.} it is the scale which simultaneously defines the boundary condition
(\ref{bcm}) and agrees with the extremal of the function
$\phi_{{\rm min}}(t)/\xi(t)$.

This procedure is illustrated in Fig.~1, where we plot
$\phi_{{\rm min}}(t)/\xi(t)$ vs. $\mu(t)$ for
$m_t=160$ GeV and supersymmetric parameters $M_S=1$
TeV, $ X_t=0,\ \c2bb=1$. The solid curve shows the two-loop result whose
stationary point satisfies eq.~(\ref{tstar}) and therefore defines the optimal
scale $t^*$. Note that there is no fine tuning in the choice
(\ref{tstar}) since any point near the stationary point would be equally
appropriate because the curve
$\phi_{{\rm min}}(t)/\xi(t)$ is very flat in that region. We have
also shown in Fig.~1 the curve $\phi_{{\rm min}}(t)/\xi(t)$ in the
one-loop approximation (dashed curve) with $\mu(t^*)$ fixed by the
two-loop result. We can see that the one-loop curve is much steeper than
the two-loop curve, which means that the one-loop result, {\em i.e.} the
tree level potential (\ref{v0}) improved by the one-loop RGE, is far
from being scale independent at any scale. This feature has also been
observed for the MSSM one-loop effective potential~\cite{gamb}.

We plot in Figs.~2a,b,c,d, $\mu(t^*)$ as a function of $m_t$ for
the different values of supersymmetric parameters $M_S=1, 10$ TeV;
$X_t^2= 0, 6 M_S^2$. In all plots the solid
curve corresponds to $\c2bb=1$ and the dashed curve to $\c2bb=0$.

Once we have determined $\mu(t^*)$ for fixed values of $m_t$ and all
supersymmetric parameters, the Higgs running mass $m_H(t)$ is given
by (\ref{mhrun}), while the mass defined as the second derivative of
the effective potential $m_{H,der}(t)$ is given by (\ref{mhder}).
By definition both masses coincide at $t^*$
\be
\label{reqd}
m_H(t^*)=m_{H,der}(t^*),
\ee
and the ratio $m_{H,der}(t)/m_H(t)$ should be equal to one
for an exactly scale independent
effective potential. Consistency of our procedure
requires the curve $m_{H,der}(t)/m_H(t)$ to be flat in the
region where we minimize, {\em i.e.} at $t^*$.
In Fig.~3 we plot $m_{H,der}(t)/m_H(t)$ for $m_t=160$ GeV and the
values of supersymmetric parameters as in Fig.~1, $M_S=1$ TeV,
$X_t=0,\ \c2bb=1$. We see that at the point $\mu(t^*)$ the curve is in
its flat region, though $\mu(t^*)$ does not exactly coincide
with the extremum of $m_{H,der}(t)/m_H(t)$.

Neither $m_H$ nor $m_t$ are physical masses. They are computed from
the effective potential, {\em i.e.} at zero external momentum, and
need to be corrected by the corresponding polarizations to obtain the
propagator pole physical masses. This will be done in section 4. For
the time being, and just to compare with other results in the
literature, we will neglect the shift to the physical poles and
define the top mass by the boundary condition (\ref{bct}) and the
Higgs mass by the usual condition
\be
\label{higgsmass}
m_H\left(\mu(t)=m_H\right)=m_H.
\ee
In this way the Higgs mass $m_H$ can be unambiguously determined. We
plot in Fig.~4 $m_H$ as a function of $m_t$ for the values of
supersymmetric parameters $M_S=1$ TeV, $X_t=0,\ \c2bb=1$. The thin
solid line corresponds to the two-loop result and the dashed-line
corresponds to the one-loop result. (We will disregard for the moment
the thick solid line, which corresponds to shifting the running
masses to propagator poles.)
The main feature that arises from Fig.~4 is that the two-loop
corrections are negative with respect to the one-loop result. This
feature is in qualitative agreement with our previous two-loop result
\cite{eq} (where the one-loop corrections to the effective potential
where not considered) and with others from different authors~\cite{hemp}.
This comparison will be done in some detail in section 5.

\nsection{Numerical Results}

We will present, in this section, the numerical results on the Higgs
mass, evaluated in the next-to-leading logarithm approximation, as a
function of the top quark mass.
The running top quark mass $m_t$ that we have been using in the
previous section was evaluated at the scale
$\mu(t)=m_t$, {\em i.e.} $m_t(m_t)=m_t$ (see eq.(\ref{bct})). However
the running mass does not coincide with
the gauge invariant pole of the top quark propagator $M_t$. In the
Landau gauge the relationship between the running $m_t$ and the
physical (pole) mass $M_t$ is given by~\cite{gray}
\be
\label{mtphys}
M_t=\left[1+{\displaystyle\frac{4}{3}}
{\displaystyle\frac{\alpha_s(M_t)}{\pi}}\right] m_t(M_t).
\ee
On the other hand, the running Higgs mass, $m_H(t)$, given by
eq.~(\ref{mhrun}), has a scale variation
\be
\label{dmh}
{\displaystyle\frac{d m_H^2(t)}{dt}}=2 \gamma m_H^2(t).
\ee
The propagator pole $M_H$ is related to the running mass through (see
Appendix B)
\be
\label{mhphys}
M_H^2=m_H^2(t) + {\rm Re}\Pi(M_H^2)-{\rm Re}\Pi(0),
\ee
where $\Pi(q^2)$ is the
(renormalized) self-energy of the Higgs boson.  In our
calculation (case (b) in section 3)
it is enough to compute the Higgs self-energies
at the one-loop level since we are computing the Higgs masses to
one-loop. The imaginary part of
$\Pi (M_H^2)-\Pi (0)$ contributes to
the Higgs width.
Assuming $M_H<2M_W$, the Higgs is stable at tree level, and
eq. (\ref{mhphys}) reads
\be
\label{remhphys}
M_H^2=m_H^2+\Pi(M_H^2)-\Pi(0).
\ee
The calculation of $\Pi(M_H^2)-\Pi(0)$ is presented in
Appendix B. We note that the scale dependence (at the one-loop level)
of $m_H(t)$, as provided by (\ref{dmh}), is cancelled by the scale
dependence of $\Pi(M_H^2)-\Pi(0)$. The only remaining scale
dependence of $M_H$ comes from two-loop contributions.
This feature is exhibited in Fig.~5 where we plot $M_H$ in the range
of $\mu(t)$ between $M_S$ and $M_Z$ for $m_t=160$ GeV and the
supersymmetric parameters $M_S=1$ TeV, $X_t=0,\ \c2bb=1$. We can see
that $M_H$ has a negligible variation ($\sim 0.5$ GeV) between $M_S$ and
$M_Z$. For the sake of comparison we have also plotted the running
mass $m_H(t)$ whose variation is more appreciable ($\sim 2.5$ GeV). This
effect is more accentuated for larger top masses.

We can see from (\ref{mtphys}) and from Fig.~4
that the effect of considering the
pole mass $M_t$ is negative on the two-loop corrections to the Higgs
mass, while the effect of $\Pi(M_H^2)-\Pi(0)$ is positive
($\sim 2$ GeV for $m_t=160$ GeV and $\sim 5$ GeV for $m_t=215$
GeV) thus partially compensating the effect of (\ref{mtphys}). The
global effect is negative and small as can be seen in Fig.~4. The
thick solid line indicates $M_H$ as a function of $M_t$. It is below
the thin solid line which was the two-loop evaluation of $m_H$ as a function
of $m_t$. The
comparison with the one-loop result (dashed line) shows that two-loop
corrections are negative with respect to the one-loop result.
Numerically they are small ($\sim 1$ GeV) for $M_t=120$ GeV and
larger ($\sim 6-7$ GeV) for $M_t=220$ GeV.

In Fig.~6a,b,c,d we plot $M_H$ as a function of $M_t$ for values of
supersymmetric parameters $M_S=1, 10$ TeV; $X_t^2= 0, 6 M_S^2;\
\c2bb=0,1$. In all cases the solid curve corresponds to $\c2bb=1$
and the dashed curve to $\c2bb=0$. Notice that the dependence on
the mixing parameter $X_t$ is sizeable.
In all the figures we have used the lower limit on the top quark
mass, $M_t> 120$ GeV at $95\%$ C.L., from the CDF dilepton channel
\cite{cdf}.
If we use the recent evidence for the top quark
production at CDF with a mass $M_t=174\pm 10^{+13}_{-12}$ GeV
\cite{cdf} and the bounds for supersymmetric parameters $M_S\leq 1$
TeV and maximal threshold correction
$X_t^2= 6 M_S^2$, we obtain the absolute upper bound
$M_H<140$ GeV.

The dependence of $M_H$ on $\tan\beta$ is exhibited in Fig.~7 where
we fix $M_t=170$ GeV and $M_S=1$ TeV. The solid curve corresponds
to the absolute upper bound for the mixing $X_t^2=6 M_S^2$ and the
dashed curve to the case of zero mixing.
Concerning the dependence of $M_H$
on the stop mixing $X_t$ parameter (or equivalently the stop splitting),
we have found it to be sizeable,
as can be seen from Figs. 6.

\nsection{Connection with other approaches}

Some papers have recently appeared aiming to estimate the mass of the
lightest Higgs boson up to next-to-leading order in the MSSM, and
with apparently contradictory results. In this section we will
comment on those papers in the context of our formalism and will exhibit
the origin of the disagreements.

\vspace{0.3cm}
\noindent In ref.~\cite{jap} the RG approach was used to estimate the mass of
the lightest Higgs boson up to next-to-leading order.
The relevant points of their calculation are the following:
i) They
considered the one-loop correction to the effective potential in the
approximation $g=g'=0$, ii) They neglected the wave-function
renormalization leading to physical masses for the top quark and
Higgs boson, iii) They minimized the effective potential at the
scale $\mu(t^*)=v$. In addition they only consider the case of zero
stop mixing, $X_t=0$.

Contrary to our results, they found that the two-loop
correction is positive
and sizeable with respect to the one-loop result. For instance in the
typical case $M_S=1$ TeV, $X_t=0, \c2bb=1$, they find for $m_t=170$ GeV the
two-loop result $m_H\sim 117$ GeV, while we find for $M_t=170$ GeV the
two-loop result $M_H\sim 111$ GeV. We have been able to trace back
the difference between the
two results to the points i)--iii) above.
In particular, the authors of
ref.~\cite{jap} find for the same values of the parameters a positive
two-loop correction with respect to the one-loop result $\sim 7$ GeV while our
two-loop result is smaller than the one-loop result by $\sim 3$ GeV. To
understand the origin of this discrepancy we have plotted in Fig.~8
the physical Higgs mass $M_H$ as a function of the minimization scale
$\mu(t^*)$ for $M_t=170$ GeV, $M_S=1$ TeV and $X_t=0$. Thick lines
correspond to $\c2bb=1$ and thin lines to $\c2bb=0$. Solid lines
represent $M_H$ evaluated in the two-loop approximation and dashed
lines in the one-loop approximation. Our chosen value of $\mu(t^*)$ is
indicated in the figure with an open diamond and
that chosen in ref. \cite{jap}, $\mu(t^*)=v$, with an
open square. We can see that $M_H$ evaluated at two-loop is very
stable against $\mu(t^*)$: it varies $\sim 5$ GeV in the whole
interval $M_Z\leq\mu(t^*)\leq M_S$ and $\sim 2$ GeV between the
diamond and the square.
However the value of $M_H$ evaluated at one-loop is very unstable. In
fact the two-loop correction changes from negative at $\mu(t)=M_Z$
$(\simgt 10\%)$ to positive at $\mu(t)=M_S$ $(\simgt 20\%)$. In the
region of our chosen value of $\mu(t^*)$ it is negative and $\sim 3$
GeV while at $\mu(t^*)=v$ it is positive and greater $\sim 7$ GeV.
Fig.~8 shows that though our choice of $\mu(t^*)$ is more in
agreement with perturbation theory than $\mu(t^*)=v$ the difference
is however not important when plotting the physical Higgs mass
(which was not considered in ref.\cite{jap}).

\vspace{0.3cm}
\noindent In ref.~\cite{hemp} the diagrammatic and effective
potential approaches were used
to evaluate the lightest Higgs mass at two-loop
order in the framework of the MSSM.
Various approximations, like $g=g'=0$ in the
effective potential, were used, and only the case of zero stop mixing,
$X_t=0$, was considered.
Our results agree
with those of ref.
\cite{hemp} within less than $\sim 3$ GeV. In fact for $M_t=150$
GeV, $X_t=0, \c2bb=1$ and $M_S=1$  TeV $(M_S=10$ TeV) ref.
\cite{hemp} finds $M_H\sim 107$ GeV $(M_H\sim 110$ GeV) while we
find from Figs.~6a and 6b $M_H\sim 103$ GeV $(M_H\sim 110$
GeV). We consider this agreement as satisfactory given the
approximations used in ref.~\cite{hemp}.
The fact that two-loop
corrections found in ref.~\cite{hemp} are sizeable with
respect to the one-loop result can be explained as a consequence of the
approximation used there to estimate the one-loop result. Taking
$M_S=10$ TeV and $M_t=150$ GeV, they found $m_H=138$ GeV at one-loop
using a simple approximation that takes into account only the leading
part ($\sim M_t^4 \log(M_S^2/M_t^2)$) of the corrections, as it is common
practice in some phenomenological analysis. Actually, a
slightly more sofisticated approximation (as the one labelled $1\beta$
in their paper) or the one-loop result obtained by a numerical integration
of the RGE (as in our approach) gives $m_H\sim 115$ GeV (both methods
agree up to a difference $\sim 2-3$ GeV due to subleading
effects, such as
those provided by considering gauge couplings
in the effective potential) for the
above mentioned values of the parameters. This has to be compared with the
two-loop result $M_H\sim 110$ GeV and shows that the net effect
of two-loop corrections is indeed negative and
small\footnote{Notice that the results of ref. \cite{hemp} only contain
one- and two-loop leading and next-to-leading contributions to the Higgs mass
while, as we have noticed before, our calculation of the Higgs mass
includes leading and next-to-leading contributions to all-loop. This
effect can be important for the case of large logarithms. We thank
R. Hempfling for a discussion on this point.}.

\vspace{0.3cm}
Finally bounds on the lightest Higgs mass have been recently analyzed
in ref.~\cite{langa} in the context of models with Yukawa
unification. The results of ref.~\cite{langa} are presented in the
one-loop approximation. They choose $M_Z$ as the minimization scale
and find larger bounds than in previous estimates. For instance it is
found, for $M_t=170$ GeV and any value of the supersymmetric parameters
such that $M_S<1$ TeV, that $M_H\simlt 102$ GeV. We have tried to
reproduce their results using our formalism. In fact, their solution
for small $\tan\beta$ is close to the fixed point solution where
$\tan\beta$ and $m_t$ are related through~\cite{barber}
\be
\label{mtfp}
m_t=(196\ {\rm GeV})[1+2(\alpha_3(M_Z)-0.12)] \sin\beta.
\ee
Now,
fixing $M_S=1$ TeV and maximal mixing\footnote{An interesting result
that can be confirmed from ref.~\cite{langa} (see Fig. 10 in that paper)
is that the maximal threshold limit is compatible with the non-existence of
color breaking minima, as we already noted. For that reason removing the
color breaking minima of the distribution plotted in Fig.~7 does not change
the upper bound on the Higgs mass.} $X_t^2=6 M_S^2$ we find,
for $M_t=170$ GeV that $\tan\beta=1.74$ from (\ref{mtfp}) and
$M_H\sim 99$ GeV if we fix $\mu(t^*)=M_Z$, in
agreement with the result of ref.~\cite{langa}. However this large
value of $M_H$ is a clear consequence of the chosen minimization
scale where two-loop corrections are very large. In fact for
$\mu(t^*)=M_Z$ our two-loop result gives $M_H\sim 85$ GeV. Moving to
the region where perturbation theory is more reliable, as we have done
along this paper, we would obtain the final two-loop result,
$M_H\sim 86$ GeV. We have plotted in Fig.~9 $M_H$ as a function of
$M_t$ for $M_S=1$ TeV and $\tan\beta$, determined from (\ref{mtfp}),
corresponding to the fixed point solution. The solid (dashed) line
corresponds to the case of maximal mixing, $X_t^2=6 M_S^2$ (zero
mixing, $X_t=0$).

\nsection{Conclusions}

We have computed in this paper the upper bounds on the mass of the
lightest Higgs boson
in the MSSM including leading and next-to-leading
logarithm radiative
corrections. We have used an RG approach by means
of a careful treatment of the effective potential in the SM,
assuming that supersymmetry is
decoupled from the SM, which we have shown to be an excellent
approximation for $M_{S}\simgt 1$ TeV or even much less.
Our results have covered the whole parameter space of
supersymmetric parameters;
in fact, when we refer
to the MSSM, we simply mean the supersymmetric version of
the SM with minimal particle content.
We have also included
QCD radiative corrections to the top quark mass,
which gives a negative
contribution to the Higgs mass as a function of the physical (pole)
top mass $M_t$, and electroweak radiative corrections to the Higgs mass
$M_H$, $\Pi(M_H^2)-\Pi(0)$, providing a positive contribution to
$M_H$ and partially cancelling the former ones. Therefore the
balance of including the contribution of QCD and electroweak
self-energies to the quark and Higgs masses is negative and small
($\simlt 2\%$ for $M_t<200$ GeV).
We also made a
reliable estimate of the total uncertainty in the final results,
which turns out to be quite small ($\sim 2\ GeV$).

Concerning the numerical results, we have found in particular that for
$M_t\leq 190$ GeV, $ M_S\leq 1$  TeV and any value of $\tan\beta$
we have $M_H< 120$ GeV for $X_t=0$, and $ M_H<140$  GeV for
$X_t^2=6 M_S^2$ (maximal threshold effect). We have also applied
our calculation to a particularly appealing scenario, namely
the assumption that the top has the Yukawa coupling in its infrared fixed
point. Then, the bounds are much more stringent. {\em E.g.}
for $M_{t}= 170$ GeV, $M_{S}\leq 1$ TeV
and $X_t^2=6 M_S^2$, we find
$M_{H}\leq 86$ GeV.

Two papers have recently tried to
incorporate radiative corrections to the Higgs mass up to the
next-to-leading order and with qualitatively different results. In
ref.~\cite{jap}, using the RG approach,
positive, and large next-to-leading corrections  with
respect to the one-loop results were found. In ref.~\cite{hemp}, using
diagrammatic and effective potential methods in a particular
MSSM as well as various
approximations, it was found that two-loop corrections are also
sizeable, but negative with respect to the one-loop result!

Using the RG approach, as ref.~\cite{jap},
we have found (unlike in ref. \cite{jap})
that two-loop corrections are negative with respect to the
one-loop result. We have traced back the origin of this
disagreement in their choice of the minimization scale.
Furthermore the authors of ref. \cite{jap} neglected various effects
(as the contribution of gauge bosons to the one--loop effective potential,
or the wave function renormalization of top quark and Higgs boson)
and considered only the case with zero stop mixing.

On the other hand, we have found that the
abnormal size of the two-loop corrections obtained in \cite{hemp}
is a consequence of an excesively rough estimate of the one-loop
result, but we are in agreement with their final two-loop result.
In fact our two-loop results differ from those of ref.
\cite{hemp} by less than $3\%$. Also our results show a large
sensitivity of the Higgs mass to the stop mixing parameter.

Finally we would like to comment briefly on the generality of our
results. As was already stated, we are assuming average
squark masses $M_S^2\gg M_Z^2$,
and that all supersymmetric particle masses are $\simgt M_S$. If
we relax the last assumption,
{\it i.e.} if some supersymmetric particles were much
lighter, the value of the quartic coupling at $M_S$ (see eq. (\ref{lms}))
would be slightly increased and, correspondingly,
our bounds would be slightly relaxed. We have made an
estimate of this effect. Assuming an extreme case where all gauginos,
higgsinos and sleptons have masses $\sim M_Z$,
we have found for $M_S=1$ TeV and
$\cos^2 2\beta=1$ an increase of the Higgs mass
$\sim 2\%$. For values of $\tan\beta$ close to one (as those appearing in
infrared fixed point scenarios) the corresponding effect is negligible.
On the other hand, our numerical results
have been computed for $M_S \geq 1$ TeV. For values of
$M_S \leq 1$ TeV the bounds on the lightest Higgs mass
are lowered. Hence, in this sense, all our
results can be considered as absolute upper bounds.

\section*{Acknowledgements}

We thank R. Hempfling and S. Peris for illuminating discussions.
We also thank M. Carena, G. Kane, M. Mangano, N. Polonsky, S. Pokorski,
A. Santamar\'{\i}a, C. Wagner and F. Zwirner for useful discussions
and comments.

\appendixA{Appendix A}

We comment here on the origin and size of threshold contributions
to the quartic coupling of the SM, $\lambda$, at the
supersymmetric scale.
We also evaluate the effective D=6 operators that are
relevant for the Higgs potential, showing that for $M_S\geq 1$ TeV
they are negligible.

As it is well known, the one-loop correction to the MSSM effective
potential is dominated by the stop contribution
\be
\label{v1mssm}
V_1^{MSSM}={\displaystyle\frac{3}{32\pi^2}}\left\{
m_1^4\left[\log{\displaystyle\frac{m_1^2}{Q^2}}
-{\displaystyle\frac{3}{2}}\right]+
m_2^4\left[\log{\displaystyle\frac{m_2^2}{Q^2}}
-{\displaystyle\frac{3}{2}}\right]
\right\}
\ee
where $m_1$, $m_2$ are the two eigenvalues of the stop mass matrix.
In a good approximation:
\be
\label{m12}
m_{1,2}^2 = M_S^2 + h_t^2 H_2 \pm h_tH_2X_t
\ee
where $M_S^2$ is the soft mass of the stops (we are assuming here
that this is the same for the left and the right stops, which is a
correct approximation since the RGE of both masses are dominated by
the same QCD term); $h_t$ is the top Yukawa coupling in the MSSM and
$X_t=A_t+\mu\cot\beta$ gives the mixing between stops ($A_t$ is the
coefficient of the soft trilinear scalar coupling involving stops and
$\mu$ is the one of the usual bilinear Higgs term in the superpotential).
Notice also that $M_S$ is basically the average of the two stop masses,
which is a reasonable choice for the supersymmetric scale when one is
dealing with the Higgs potential, as it is our case. Thus we identify
the scale $Q$ at which we are evaluating the threshold effects with $M_S$.

Now it is straightforward to obtain from (\ref{v1mssm}) the contribution
to the $H^4$ operator (recall that $H$ is the SM Higgs doublet given by
eq.(\ref{hsm})). That is
\be
\label{H4}
{\displaystyle\frac{3}{32\pi^2}}\left\{
\left( \frac{2X_t^2}{M_S^2}\right)h_t^4\ -
\ \left( \frac{X_t^4}{6M_S^2}\right)h_t^4\right\}\ H^4\ \ ,
\ee
where have redefined $h_t$ to be the usual top Yukawa coupling in the SM.
This gives a threshold contribution to the SM quartic coupling
\be
\label{deltal2}
\Delta\lambda={\displaystyle\frac{3 h_t^4}{16
\pi^2}}{\displaystyle\frac{X_t^2}{M_S^2}}\left(2-
{\displaystyle\frac{X_t^2}{6 M_S^2}}\right).
\ee

Alternatively, eq. (\ref{deltal2}) can be obtained diagrammatically
from two kinds of diagrams exchanging stops (see
the first paper of ref.~\cite{rg}).
Both methods give the same result.

It is worth noticing that (\ref{deltal2}) has a maximum
for $X_t^2=6M_S^2$, which means that the threshold effect cannot be
arbitrarily large. Note also that $X_t^2=6M_S^2$
is barely consistent with the bound from color conserving
minimum~\cite{color}, so the case of maximal threshold represents
an extreme situation.

Analogously, we can obtain from (\ref{v1mssm}) the relevant effective
D=6 operators, {\em i.e.} those proportional to $H^6$. These turn out to be
\be
\label{H6}
{\displaystyle\frac{3}{32\pi^2}}\left\{
\left( \frac{2}{3M_S^2}\right)\ -
\left( \frac{X_t^2}{M_S^4}\right)\ +
\left( \frac{X_t^4}{3M_S^6}\right)\ -
\ \left( \frac{X_t^6}{30M_S^8}\right)\right\}\ h_t^6H^6\ \ ,
\ee
which could also be obtained diagrammatically
from four kinds of diagrams exchanging stops. It is easy
to see that (\ref{H6}) produces negligible modifications
in the process of electroweak breaking and in the Higgs mass.
For example, for $M_S=1$ TeV and maximal threshold in (\ref{deltal2}),
{\em i.e.} $X_t^2=6M_S^2$, (\ref{H6}) gives modifications in the
Higgs mass suppressed by a factor $\sim 1/150$ with respect
to those induced by (\ref{deltal2}). This also means that, regarding
the Higgs potential, it is safe to decouple the MSSM from the SM
for $M_S=1$ TeV  or even substantially smaller.

\appendixB{Appendix B}

In this Appendix we shall discuss in more detail the relation
between the running mass of the Higgs boson,
extracted from the effective potential, and the physical Higgs mass
and give the complete expression for the latter.

The physical mass of the Higgs boson field $M_H$ is defined as the
pole of the propagator and it is both renormalization
scheme and  gauge (if calculated at all orders of perturbation)
independent.

We start with the Lagrangian
\begin{eqnarray}
\label{lagr}
{\cal L}&=&\frac{1}{2}\left(\partial\phi_0\right)^2
-\frac{1}{2}m_0^2\phi_0^2+\ldots \nonumber\\
&=&
\frac{1}{2}\left(\partial\phi_R\right)^2
-\frac{1}{2}m_R^2\phi_R^2
+\frac{1}{2}\delta Z_H\left(\partial\phi_R\right)^2
-\frac{1}{2}\delta m^2\phi_R^2+\ldots \ \ ,
\end{eqnarray}
where $\phi_0$ and  $\phi_R=Z_H^{-1/2}\phi_0=(1+\delta Z_H)^{-1/2}\phi_0$
are the bare and renormalized Higgs fields, and $m_0$ and $m_R$ are
the bare and renormalized masses. With this convention
$m_0^2=m_R^2+\delta m^2-\delta Z_H m_R^2$. Denoting by
$\Gamma_{0}$ ($\Gamma_{R}$) the inverse of the one-loop corrected
bare (renormalized) propagator, we have
\begin{eqnarray}
\label{gg}
\Gamma_{R}(p^2)&=&Z_H\Gamma_{0}(p^2)=Z_H \left[
p^2-m_R^2-\delta m^2+\delta Z_H m_R^2-\Pi_0(p^2) \right]\nonumber\\
&=&
p^2+\delta Z_H p^2-\left(m_R^2+\delta m^2\right)-\Pi_0(p^2)\ \ ,
\end{eqnarray}
where $\Pi_0(p^2)$ is the (unrenormalized) self-energy of the Higgs
boson field and in the last equality we have neglected higher order
terms in the perturbative expansion. The renormalized self-energy
is defined as
\begin{equation}
\label{renpi}
\Pi_R(p^2)=\Pi_0(p^2)+\delta m^2-\delta Z_H p^2.
\end{equation}
Thus
\begin{equation}
\label{gr}
\Gamma_{R}(p^2)=
p^2-\left(m_R^2+\Pi_R(p^2) \right)\ \ .
\end{equation}
Consequently, the physical (pole) mass, $M_H^2$, satisfies the relation
\begin{equation}
\label{MHphys}
M_H^2=m_R^2+\Pi_R(p^2=M_H^2)\ \ .
\end{equation}
On the other hand, the running mass $m_H^2$, defined as the
second derivative of the renormalized effective potential
(see eqs. (\ref{mhstar}) and (\ref{mhrun})), is given by
\begin{equation}
\label{mhmh}
m_H^2=\frac{\partial^2 V_{eff}}{\partial \phi^2}=
-\Gamma_{R}(p^2=0)=
m_R^2+\Pi_R(p^2=0) \ \ .
\end{equation}
Comparing (\ref{MHphys}) and (\ref{mhmh}) we have
\begin{equation}
\label{MHdelta}
M_H^2= m_{H}^2 +\Delta \Pi,
\end{equation}
where we have defined (we drop the subscript R from $\Pi_R$)
\begin{equation}
\label{deltadepi}
\Delta \Pi\equiv\Pi (p^2=M_H^2)-\Pi (p^2=0).
\end{equation}
Note that $m_H^2$ defined in (\ref{mhmh}) is renormalization
scheme dependent, as $V_{eff}$ is, while $M_H^2$ is not. In
particular, in eq.(\ref{MHdelta}) both $m_{H}^2$ and $\Delta \Pi$
depend on the renormalization scale $\mu$ in such a
way that $M_H^2$ results to be scale
independent (at least to ${\cal O}(\hbar)$).
In fact, from eq. (\ref{renpi}) and the definition
\begin{equation}
\gamma=\frac{1}{2}\frac{d\log Z_H}{d\log\mu}
\end{equation}
we easily obtain
\begin{equation}
\frac{d \Delta \Pi}{d\log\mu}=-2\gamma M_H^2
\end{equation}
which cancels the scale dependence of $m_H^2$ (see eq.(\ref{dmh})).

We now want to give the complete expression for the physical mass
$M_H^2$. The quantity $\Delta \Pi$ in the Landau gauge
is given by the sum of the following terms
\begin{eqnarray}
\label{B8}
\Delta \Pi&=&\Delta \Pi_{tt}
\:\:\:\: ({\rm top}\:\:{\rm contribution})\nonumber\\
&+&
\Delta \Pi_{W^{\pm}W^{\mp}}+\Delta \Pi_{Z^{0}Z^{0}}\nonumber\\
&+&\Delta \Pi_{W^{\pm}\chi^{\mp}}+
\Delta \Pi_{Z^{0}\chi_{3}}
\:\:\:\: (\mbox{gauge and Goldstone bosons contribution})\nonumber\\
&+& \Delta \Pi_{\chi^{\pm}\chi^{\mp}}+
\Delta \Pi_{\chi_{3}\chi_{3}}\nonumber\\
&+&\Delta \Pi_{HH}\:\:\:\: (\mbox{pure scalar bosons
contribution}).
\end{eqnarray}
In eq.~(\ref{B8}) we have
taken into account only the contribution from the heaviest fermion, the top,
and indicated by $\chi^{\pm}$ and $\chi_{3}$ the charged and the neutral
Goldstone bosons, respectively.

The complete expression for the different
contributions to $\Delta \Pi$ calculated in the $\overline{MS}$ scheme is, for
$M_H<2 M_W$:

$i)$ Top contribution:
\begin{equation}
\Delta \Pi_{tt}=\frac{3h_{t}^{2}}{8\pi^2}
\left\{-2M_{t}^2\left[Z\left(\frac{M_{t}^2}{M_H^2}\right)
-2\right]+\frac{1}{2}M_H^2\left[\log
\frac{M_{t}^2}{\mu^2}+
Z\left(\frac{M_{t}^2}{M_H^2}\right)
-2\right]\right\}.
\end{equation}

$ii)$ Gauge bosons and Goldstone bosons contribution:
\begin{eqnarray}
&&\Delta \Pi_{W^{\pm}W^{\mp}}+\Delta \Pi
_{Z^{0}Z^{0}} +\Delta \Pi_{W^{\pm}\chi^{\mp}}+
\Delta \Pi_{Z^{0}\chi_{3}}\nonumber\\
&=&\frac{g^2 M_{W}^2}{8\pi^2}
\left(I_{W}^{(1)}+I_{W}^{(2)}+\frac{M_H^2}{2}
I_{W}^{(3)}-M_H^2
I_{W}^{(4)}\right.\nonumber\\
&+& \left.\frac{M_H^2}{4}
I_{W}^{(5)}-\frac{M_H^4}{2}
I_{W}^{(6)}\right)-\frac{1}{2}\frac{g^2 M_H^2}{16\pi^2}
\left(
I_{W}^{(7)}(\mu^2)+I_{W}^{(8)}(\mu^2)\right)\nonumber\\
&+&\frac{1}{2}\frac{g^2 M_H^4}{16\pi^2}
I_{W}^{(9)}+\frac{1}{2}
\left\{ \begin{array}{c}
M_{W}\rightarrow M_{Z}\cr\\
g^2\rightarrow g^2+g^{\prime 2}
\end{array}
\right\},
\end{eqnarray}
where all the masses in the above expression have to be understood
as the physical ones.
The $I_{W}^{i}$ ($i=1,....,9$) functions read
\begin{flushleft}
\begin{eqnarray}
I_{W}^{1}&=&\int_0^1 dx \log\left[ 1 - \frac{M_H^2}{M_W^2}x(1-x)- i\epsilon
\right]\;=\;Z\left(\frac{M_{W}^2}{M_{H}^2}\right)-2,\nonumber\\
 & \nonumber \\
I_{W}^{2}&=&\int_0^1\int_0^1 dx dy\; y \log\left[1-\frac{M_H^2}{M_W^2}
\frac{y(1-y)}{(1-xy)}- i\epsilon
\right]\;=\; -\frac{11}{12}-\frac{1}{6}
\frac{M_{W}^2}{M_{H}^2}\nonumber
\\
&&+\frac{1}{6}\left(4-
\frac{M_{H}^2}{M_{W}^2}\right)
Z\left(\frac{M_{W}^2}{M_{H}^2}\right)
+\frac{1}{6}\frac{\left(M_{H}^2-M_{W}^2\right)^3}
{M_{H}^4 M_{W}^2}\log\left(1-
\frac{M_{H}^2}{M_{W}^2}\right),\nonumber\\
 & \nonumber \\
I_{W}^{3}&=&\int_0^1\int_0^1 dx dy \frac{y^3}{
M_W^2(1-xy)-M_H^2 y(1-y) - i\epsilon}
\;=\;\frac{5}{6}\frac{1}{M_{H}^2}+\frac{1}{3}
\frac{M_{W}^2}{M_{H}^4}\nonumber\\
&&+\frac{1}{3 M_{H}^2}
\left(
\frac{M_{H}^2}{M_{W}^2}-1\right)
Z\left(\frac{M_{W}^2}{M_{H}^2}\right)+
\frac{1}{3 M_{W}^2}
\left(
\frac{M_{W}^6}{M_{H}^6}-1\right)
\log\left(1-
\frac{M_{H}^2}{M_{W}^2}\right),\nonumber\\
 & \nonumber \\
I_{W}^{4}&=&\int_0^1\int_0^1 dx dy \frac{y^2}{
M_W^2(1-xy)-M_H^2 y(1-y) - i\epsilon}\nonumber\\
&=&\frac{1}{2}\frac{1}{M_{H}^2}
+\frac{1}{2M_{W}^2}Z\left(\frac{M_{W}^2}{M_{H}^2}\right)
+\frac{1}{2 M_{W}^2}
\left(
\frac{M_{W}^4}{M_{H}^4}-1\right)
\log\left(1-
\frac{M_{H}^2}{M_{W}^2}\right),\nonumber\\
 & \nonumber \\
I_{W}^{5}&=&\int_0^1\int_0^1\int_0^1 dx dy dz \frac{z (1-z)}{
M_W^2(1-y-z(x-y))-M_H^2 z(1-z) - i\epsilon}\nonumber\\
&=&-\frac{1}{3}\frac{1}{M_{H}^2}
+\frac{1}{6M_{W}^2}\left(4-
\frac{M_{H}^2}{M_{W}^2}\right)
Z\left(\frac{M_{W}^2}{M_{H}^2}\right)\nonumber
\\
&&+\frac{\left(M_{H}^2-M_{W}^2\right)^3}
{3M_{H}^4 M_{W}^4}\log\left(1-
\frac{M_{H}^2}{M_{W}^2}\right)
-\frac{1}{6}\frac{M_{H}^2}{M_{W}^4}
\log
\frac{M_{H}^2}{M_{W}^2}+\frac{i\pi}{6}\frac{M_H^2}{M_W^4},\nonumber\\
 & \nonumber \\
I_{W}^{6}&=&\int_0^1\int_0^1\int_0^1 dx dy dz \frac{z^3 (1-z)}{\left[
M_W^2(1-y-z(x-y))-M_H^2 z(1-z) - i\epsilon\right]^2}\nonumber\\
&=&-\frac{2}{3}\frac{1}{M_{H}^4}
+\frac{1}{3M_{W}^2 M_{H}^2}\left(1-
\frac{M_{H}^2}{M_{W}^2}\right)
Z\left(\frac{M_{W}^2}{M_{H}^2}\right)\nonumber
\\
&&+
\frac{\left(M_{H}^2-M_{W}^2\right)^3}
{3M_{H}^6 M_{W}^4}\log\left(1-
\frac{M_{H}^2}{M_{W}^2}\right)
+\frac{1}{3M_{W}^4}
\log\left(1-\frac{M_{H}^2}{M_{W}^2}\right)
\nonumber
\\
&&-\frac{1}{3}\frac{M_{W}^2}{M_{H}^6}
\log\left(1-
\frac{M_{H}^2}{M_{W}^2}\right)+\frac{1}{3M_W^4}\log
\frac{M_{W}^2}{M_{H}^2}+\frac{i\pi}{3}\frac{1}{M_W^4},\nonumber\\
 & \nonumber \\
I_{W}^{7}&=&\int_0^1 dx (1+2x)\log\left[ \frac{M_W^2x-M_H^2x(1-x)- i\epsilon}
{\mu^2}
\right]\nonumber\\
&=& -4 +2\log
\frac{M_{W}^2}{\mu^2}+\frac{M_{W}^2}{M_{H}^2}
+\left(\frac{M_{W}^4}{M_{H}^4}-3\frac{M_{W}^2}{M_{H}^2}
+2\right)
\log\left(1-
\frac{M_{H}^2}{M_{W}^2}\right),\nonumber\\
 & \nonumber \\
I_{W}^{8}&=&\int_0^1 dx \log\left[\frac{M_W^2 x - i
\epsilon}{\mu^2}\right]\;=\;
 -1+\log
\frac{M_{W}^2}{\mu^2},\nonumber\\
 & \nonumber \\
I_{W}^{9}&=&\int_0^1\int_0^1 dx dy \frac{1}{M_W^2x-M_H^2y-i\epsilon}\nonumber\\
&=&
\left(\frac{1}{M_{W}^2}-\frac{1}
{M_{H}^2}\right)
\log\left(1-\frac{M_{H}^2}{M_{W}^2}\right)
+\frac{1}{M_{W}^2}\log\frac{M_W^2}{M_H^2}+\frac{i\pi}{M_W^2}
,
\end{eqnarray}
\end{flushleft}
where $Z(x)$ is the function
\begin{eqnarray}
Z(x)&=&\left\{
\begin{array}{ll}
2 A \:{\rm tan}^{-1}(1/A), & \mbox{if $x>1/4$}\\
A \:\log\left[(1+A)/(1-A)\right],&\mbox{if $x<1/4$}
\end{array}\right.\nonumber\\
A&\equiv&|1-4x|^{1/2}.
\end{eqnarray}

The terms containing the factor $\log(1-M_H^2/M_{W,Z}^2)$ develop an
imaginary part in the region $M_{W,Z}<M_H<2M_W$ corresponding to the
unphysical decays $H\rightarrow W^{\pm}\chi^{\mp},Z^0\chi^3$. However,
this imaginary part, along with the whole factor $\log(1-M_H^2/M_{W,Z}^2)$,
cancels in (B.13). In fact, using (B.14), (B.13) can be written as:
\begin{eqnarray}
&&\Delta \Pi_{W^{\pm}W^{\mp}}+\Delta \Pi
_{Z^{0}Z^{0}} +\Delta \Pi_{W^{\pm}\chi^{\mp}}+
\Delta \Pi_{Z^{0}\chi_{3}}\nonumber\\
&=&\frac{g^2 M_{W}^2}{8\pi^2}
\left[-3 +\frac{5}{4}\frac{M_H^2}{M_W^2}
+\frac{1}{2}\left(3-\frac{M_H^2}{M_W^2}+\frac{M_H^4}{4 M_W^4}\right)
Z\left(\frac{M_W^2}{M_H^2}\right)
-\frac{M_H^4}{8 M_W^4}\log\frac{M_H^2}{M_W^2}\right.
\nonumber\\
&-&\left.\frac{3M_H^2}{4M_W^2}\log\frac{M_W^2}{\mu^2}+\frac{i\pi}{8}
\frac{M_H^2}{M_W^2}
\right]+\frac{1}{2}
\left\{ \begin{array}{c}
M_{W}\rightarrow M_{Z}\cr\\
g^2\rightarrow g^2+g^{\prime 2}
\end{array}
\right\}.
\end{eqnarray}
There is also an explicit imaginary part in (B.16) giving rise to
\be
{\rm Im}(\Delta \Pi_{W^{\pm}W^{\mp}}+\Delta \Pi
_{Z^{0}Z^{0}} +\Delta \Pi_{W^{\pm}\chi^{\mp}}+
\Delta \Pi_{Z^{0}\chi_{3}})=\frac{3 g^2}{128\pi}\frac{M_H^4}{M_W^2}
\ee
which will also cancel, as we will see.

$iii)$ Pure scalar bosons contribution:
the contribution to $\Delta \Pi$
 coming from the pure scalar sector deserves more
attention and we want to discuss it in more details.

It is well-known that in the Landau gauge
the Goldstone bosons $\chi$'s do have a field dependent
mass $m_{\chi}(\phi)=-m^2+\lambda \phi^2/2$ which vanishes at the minimum
of the potential $V_{{\rm eff}}(\phi)$. As a consequence, the running mass
$m_{H}$ presents an infrared logarithmic divergence when Goldstone bosons
are included in the effective potential $V_{{\rm eff}}(\phi)$. On the other
hand, the physical mass $M_H$ must be finite and gauge independent,
so the divergent contribution coming from the Goldstone bosons to
$m_{H}^2$ must be cancelled by an equal (and opposite in sign) contribution
of the same excitations to $\Delta \Pi$. To see it explicitly,
one can imagine the Goldstone bosons to have a fictitious mass
$m_{\chi}$ and calculate their contribution $\Delta
m_{H}^{2}$
to the running square mass
$m^2_{H}$.
It is not difficult to see that this contribution
from the effective potential is
\begin{equation}
\label{B13}
\Delta m_{H}^{2}=\frac{3}{128\pi^2}\frac{g^2 M_H^4}{M_{W}^2}\:
\left[3\log\frac{M_H^2}{\mu^2(t^{\star})}+
\log\frac{m_{\chi}^2}{\mu^2(t^{\star})}\right]+
{\cal O}(\hbar^2),
\end{equation}
where the scale $\mu(t^{\star})$ is defined in the text.
On the other hand,
the contribution to $\Delta \Pi$ from the pure scalar sector
reads
\begin{equation}
\Delta \Pi_{HH}+
\Delta \Pi_{\chi^{\pm}\chi^{\mp}}+
\Delta \Pi_{\chi_{3}\chi_{3}}=
\frac{3}{128\pi^2}\frac{g^2 M_H^4}{M_{W}^2}\left[\pi\sqrt{3}
-8+ Z\left(\frac{m_{\chi}^2}{M_H^2}\right)-i\pi\right],
\end{equation}
where the last term comes from the Feynman diagrams involving Goldstone
bosons. The explicit imaginary part in (B.19) would correspond to the
unphysical decays $H\rightarrow \chi^{\pm}\chi^{\mp},\chi^3\chi^3$ and
cancels against eq. (B.17). Expanding now the function
$Z\left(m_{\chi}^2/M_H^2\right)$ around $m_{\chi}^2=0$ one can easily
show that the logarithmic divergence in $\Delta
m_{H}^{2}$ disappears and the final result for the pure
scalar bosons contribution to $\Delta M_H^2$ is finite and given by
\begin{equation}
\label{B15}
\Delta M_{H}^{2}=
\frac{3}{128\pi^2}\frac{g^2 M_H^4}{M_{W}^2}\left[\pi\sqrt{3}
-8+4\log\frac{M_H^2}{\mu^2(t^{\star})}\right].
\end{equation}
Finally it is worth making a couple of comments. First, we have included
the pure scalar bosons sector in $\Delta \Pi$ for the sake of
completeness, but now one is no longer allowed to compare the physical mass
given in eq.~(\ref{MHdelta}) with the running mass since the latter
(see eq. (\ref{B13})) is not well defined
when Goldstone bosons are taken into account. Secondly,
there is no scale dependence in the last expression (\ref{B15})
(the scale $\mu(t^{\star})$ is fixed (see the text)),
in agreement with the fact that the
$\lambda$-dependence of the anomalous dimension
$\gamma$ of the Higgs field arises at two-loop.

\def\MPL #1 #2 #3 {{\em Mod.~Phys.~Lett.}~{\bf#1}\ (#2) #3 }
\def\NPB #1 #2 #3 {{\em Nucl.~Phys.}~{\bf B#1}\ (#2) #3 }
\def\PLB #1 #2 #3 {{\em Phys.~Lett.}~{\bf B#1}\ (#2) #3 }
\def\PR #1 #2 #3 {{\em Phys.~Rep.}~{\bf#1}\ (#2) #3 }
\def\PRD #1 #2 #3 {{\em Phys.~Rev.}~{\bf D#1}\ (#2) #3 }
\def\PRL #1 #2 #3 {{\em Phys.~Rev.~Lett.}~{\bf#1}\ (#2) #3 }
\def\PTP #1 #2 #3 {{\em Prog.~Theor.~Phys.}~{\bf#1}\ (#2) #3 }
\def\RMP #1 #2 #3 {{\em Rev.~Mod.~Phys.}~{\bf#1}\ (#2) #3 }
\def\ZPC #1 #2 #3 {{\em Z.~Phys.}~{\bf C#1}\ (#2) #3 }

\newpage
\section*{Figure Captions}
\begin{description}
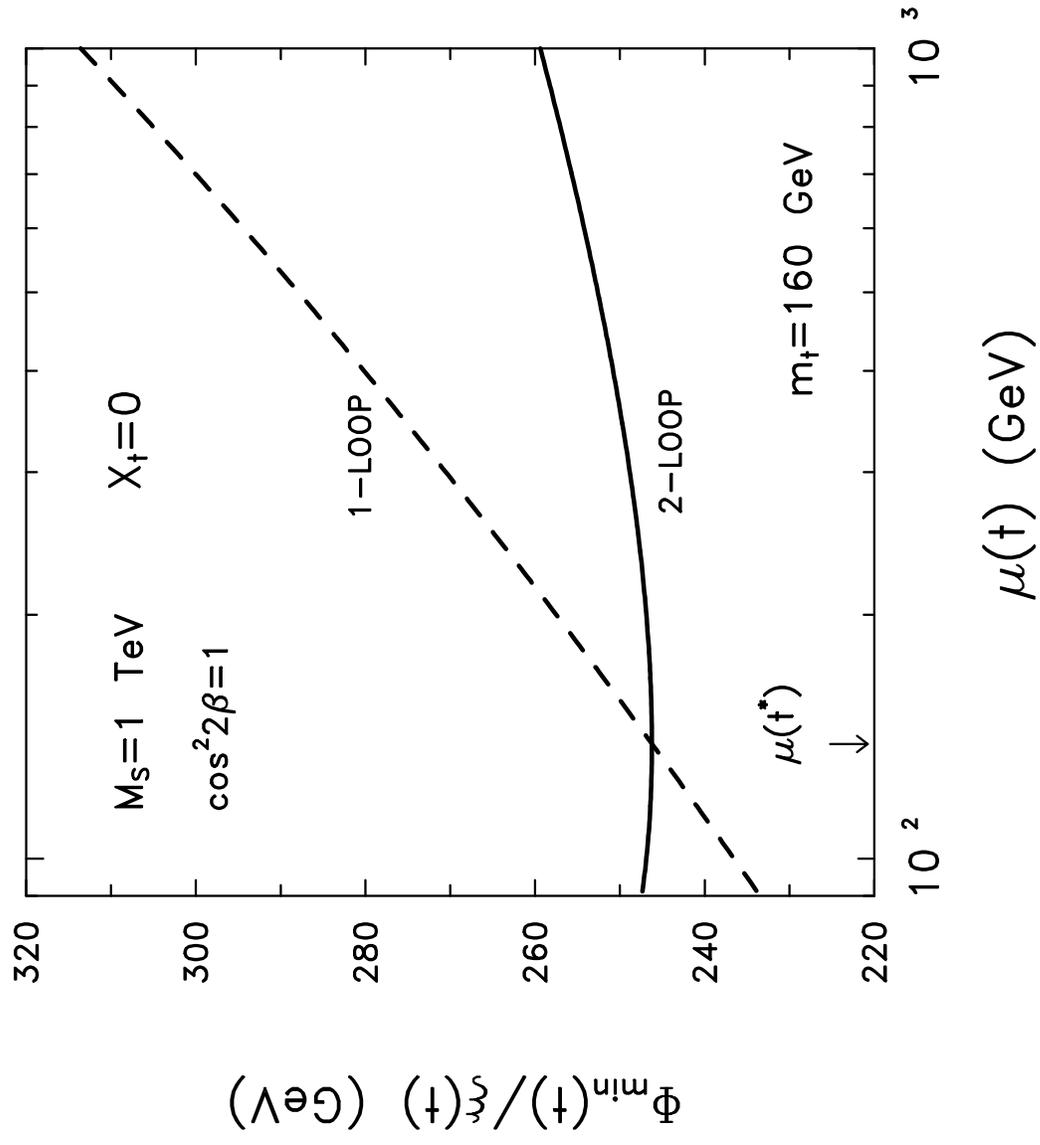
\item[Fig.~1] Plot of $\phi_{\rm min}(t)/\xi(t)$ as a function of
$\mu(t)$ for $m_t=160$ GeV and supersymmetric parameters: $M_S=1$
TeV, $X_t=0, \c2bb=1$. The solid (dashed) curve corresponds to the two
loop (one-loop) approximation.

\item[Fig.~2] Plot of $\mu(t^*)$ as a function of $m_t$ for $\c2bb=1$
(solid line) and $\c2bb=0$ (dashed line) and values of supersymmetric
parameters: a) $M_S=1$ TeV, $X_t=0$; b) $M_S=10$ TeV, $X_t=0$; c)
$M_S=1$ TeV, $X_t^2=6 M_S^2$; and d) $M_S=10$ TeV, $X_t^2=6M_S^2$.

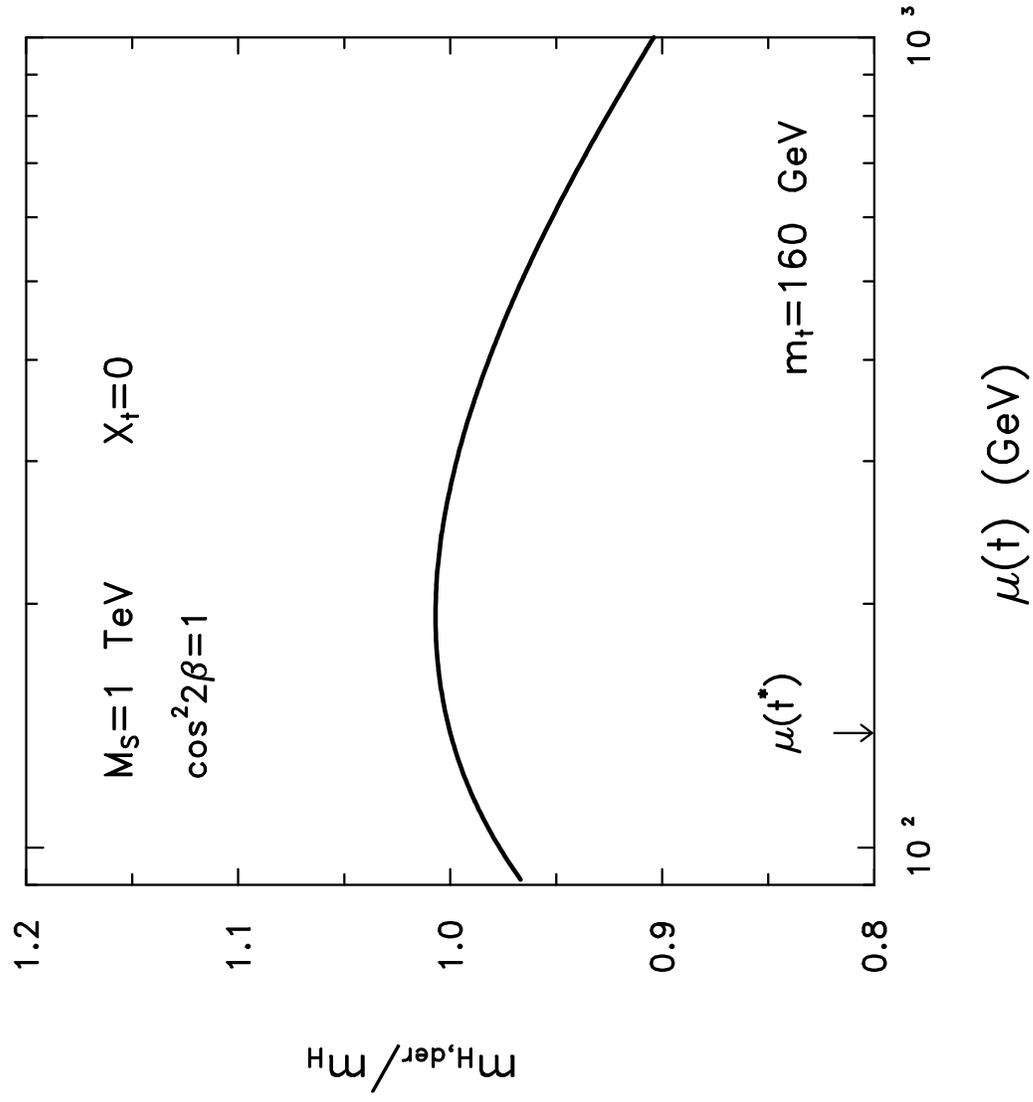
\item[Fig.~3] Plot of $m_{H,der}(t)/m_H(t)$ as a function of $\mu(t)$
for $m_t=160$ GeV and the supersymmetric parameters: $M_S=1$
TeV, $X_t=0, \c2bb=1$.

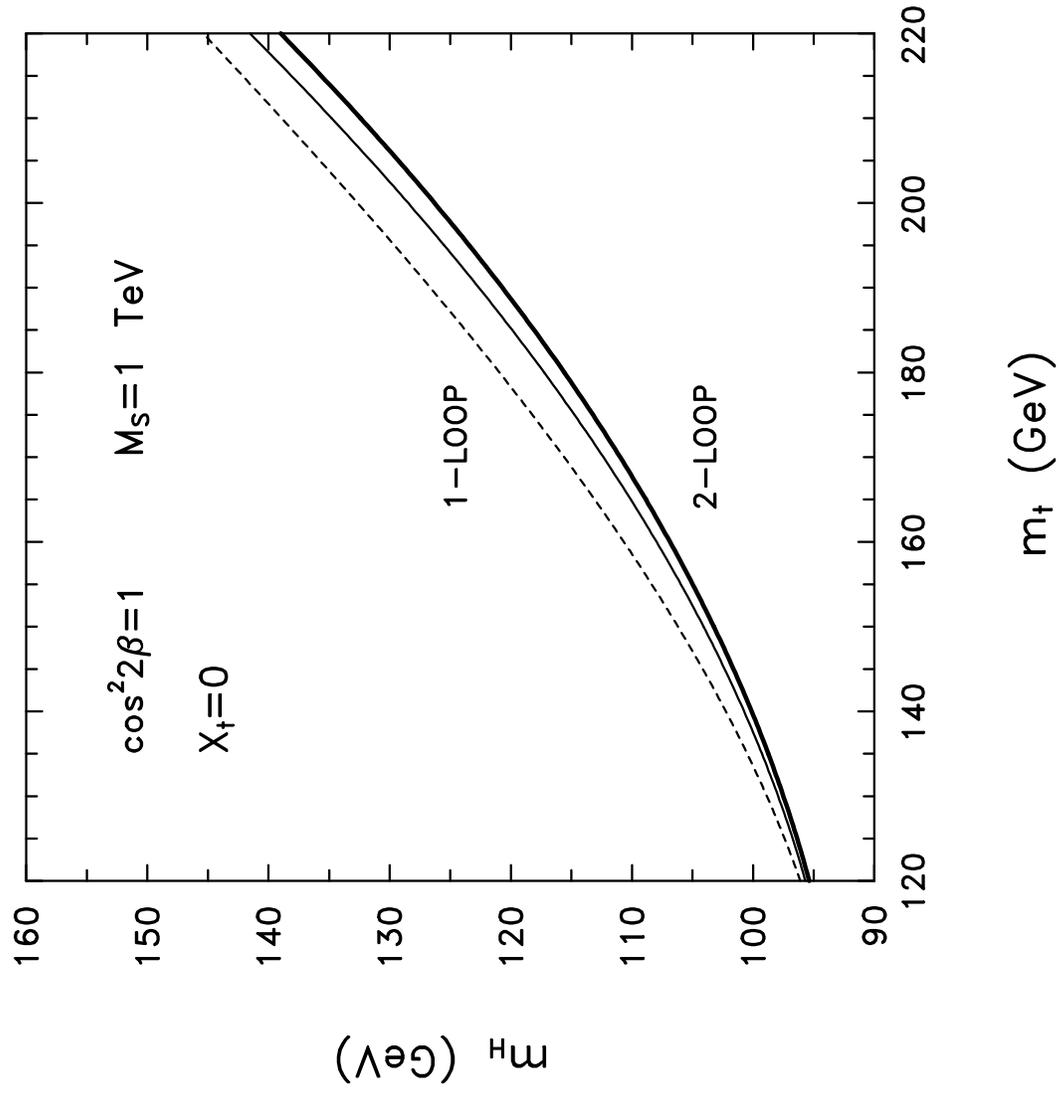
\item[Fig.~4] The thin lines correspond to $m_H$ as a function of
$m_t$ in the two-loop (solid) and one-loop (dashed) approximation for
supersymmetric parameters: $M_S=1$
TeV, $X_t=0, \c2bb=1$. The thick solid line is the plot of $M_H$, (4.3),
as a function of $M_t$, (4.1), in the two-loop approximation and for
the same values of the supersymmetric parameters.

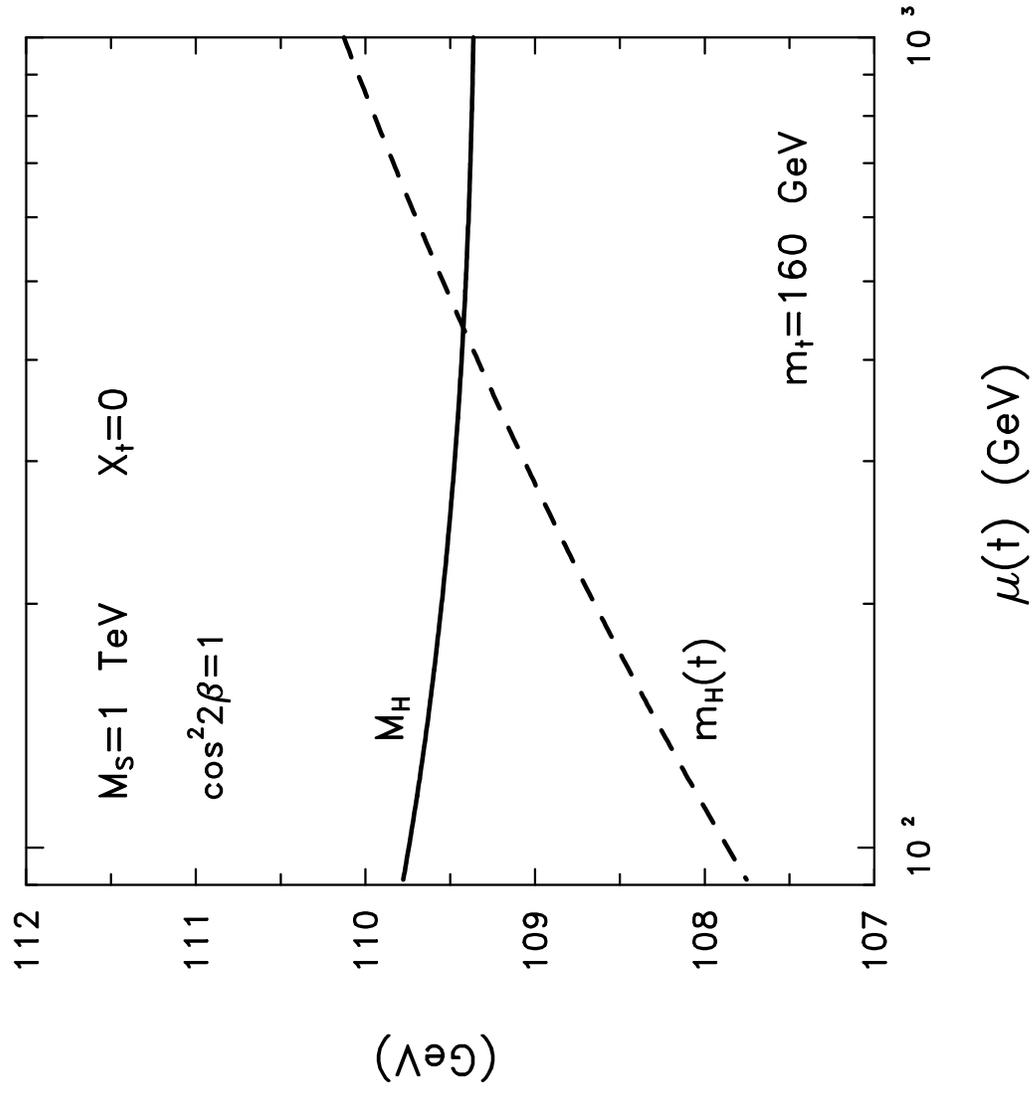
\item[Fig.~5] Plot of the physical, $M_H$ (solid curve), and running,
$m_H$ (dashed curve) Higgs mass as a function of the scale $\mu(t)$
for supersymmetric parameters $M_S=1$  TeV, $X_t$$=$$0$, $\c2bb=1$.

\item[Fig.~6] Plot of $M_H$ as a function of $M_t$ for $\c2bb=1$
(solid line) and $\c2bb=0$ (dashed line) and values of supersymmetric
parameters: a) $M_S=1$  TeV, $X_t=0$; b) $M_S=10$ TeV, $X_t=0$; c)
$M_S=1$ TeV, $X_t^2=6 M_S^2$; and d) $M_S=10$ TeV, $X_t^2=6 M_S^2$.

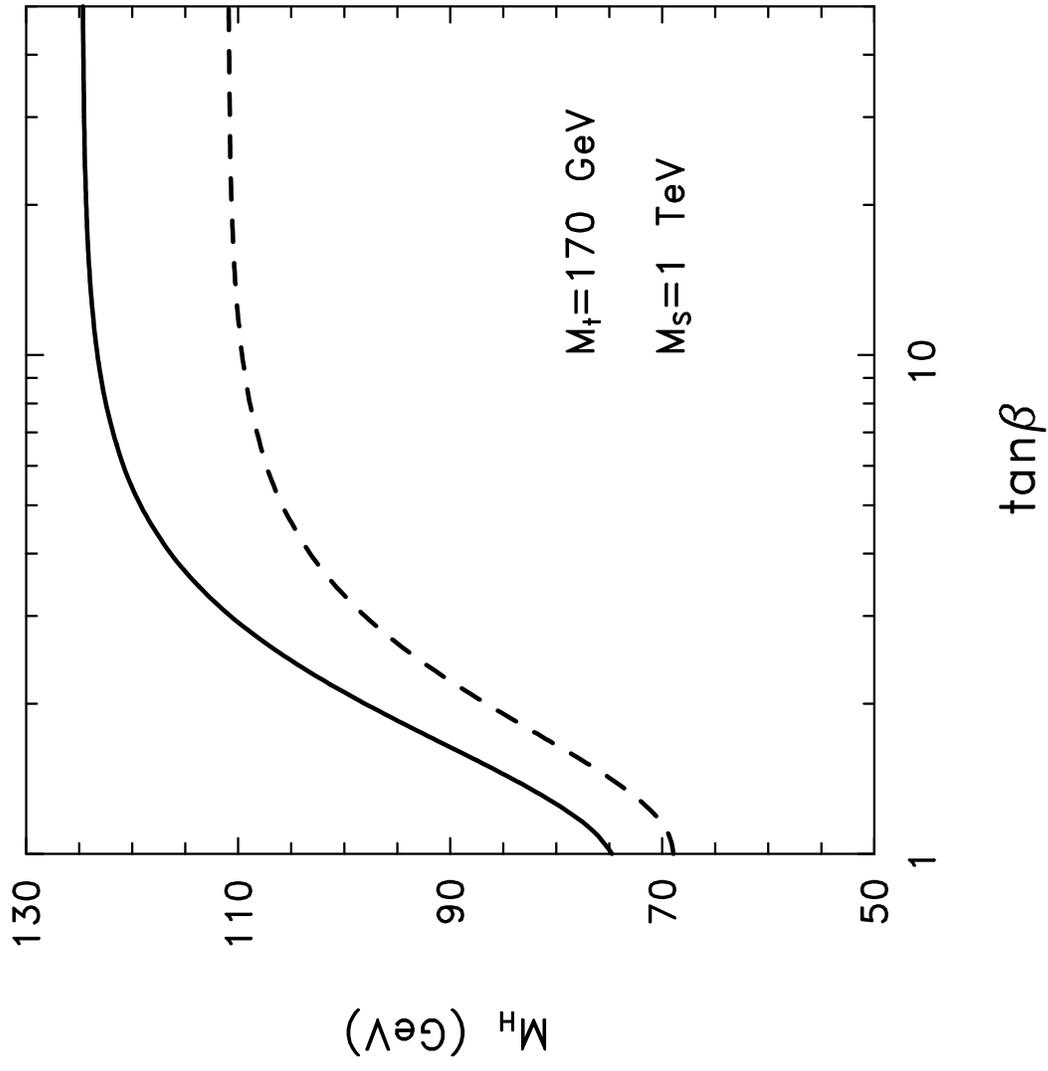
\item[Fig.~7] Plot of $M_H$ as a function of $\tan\beta$ for
$M_t=170$ GeV and $M_S=1$ TeV. The solid (dashed) curve corresponds
to the $X_t^2=6M_S^2$ ($X_t=0$) case.

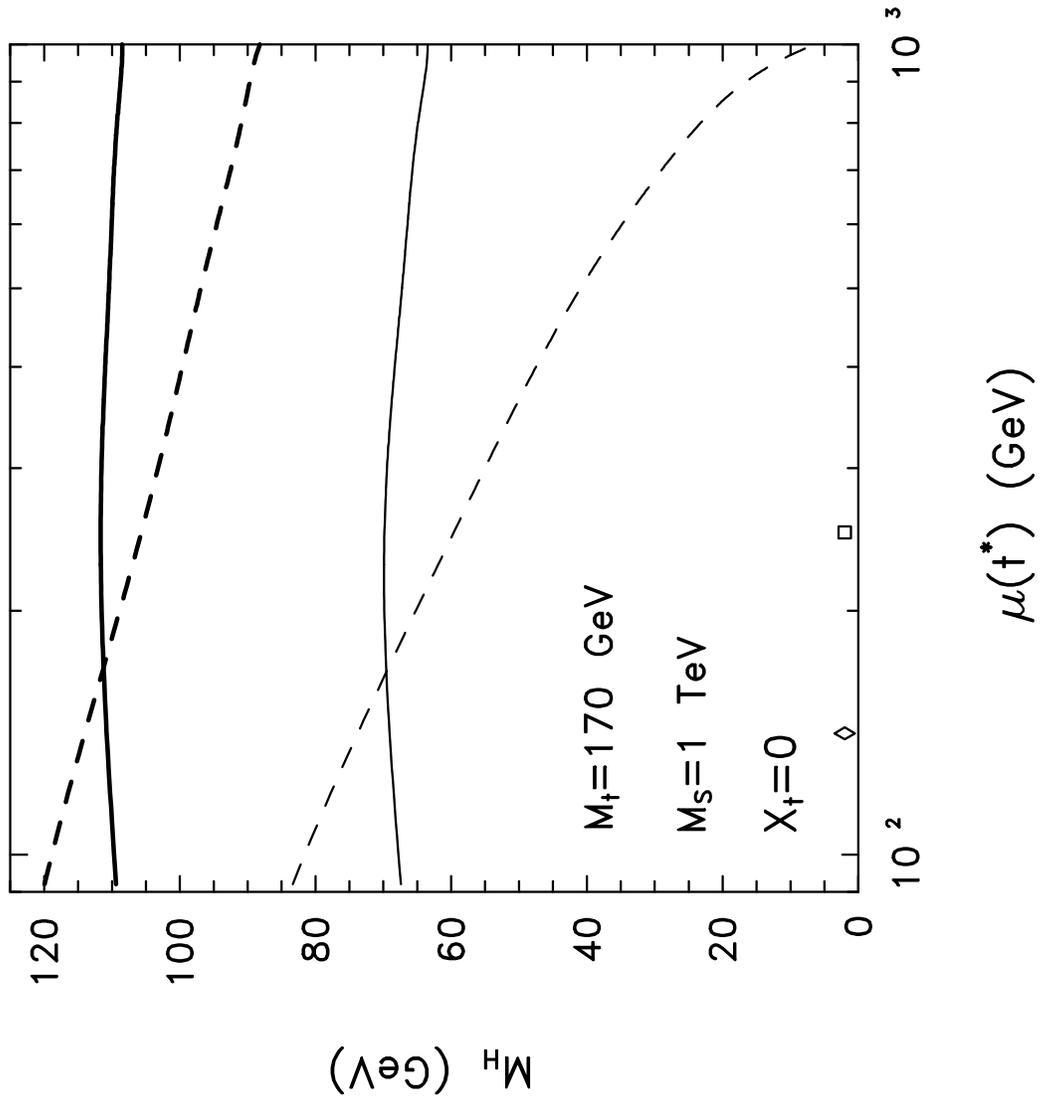
\item[Fig.~8] Plot of $M_H$ as a function of $\mu(t^*)$ for $M_t=170$
GeV, $M_S=1$ TeV, $X_t=0$. Solid (dashed) lines are the two-loop (one-loop)
approximation, thick (thin) lines correspond to $\c2bb=1$ ($\c2bb=0$).

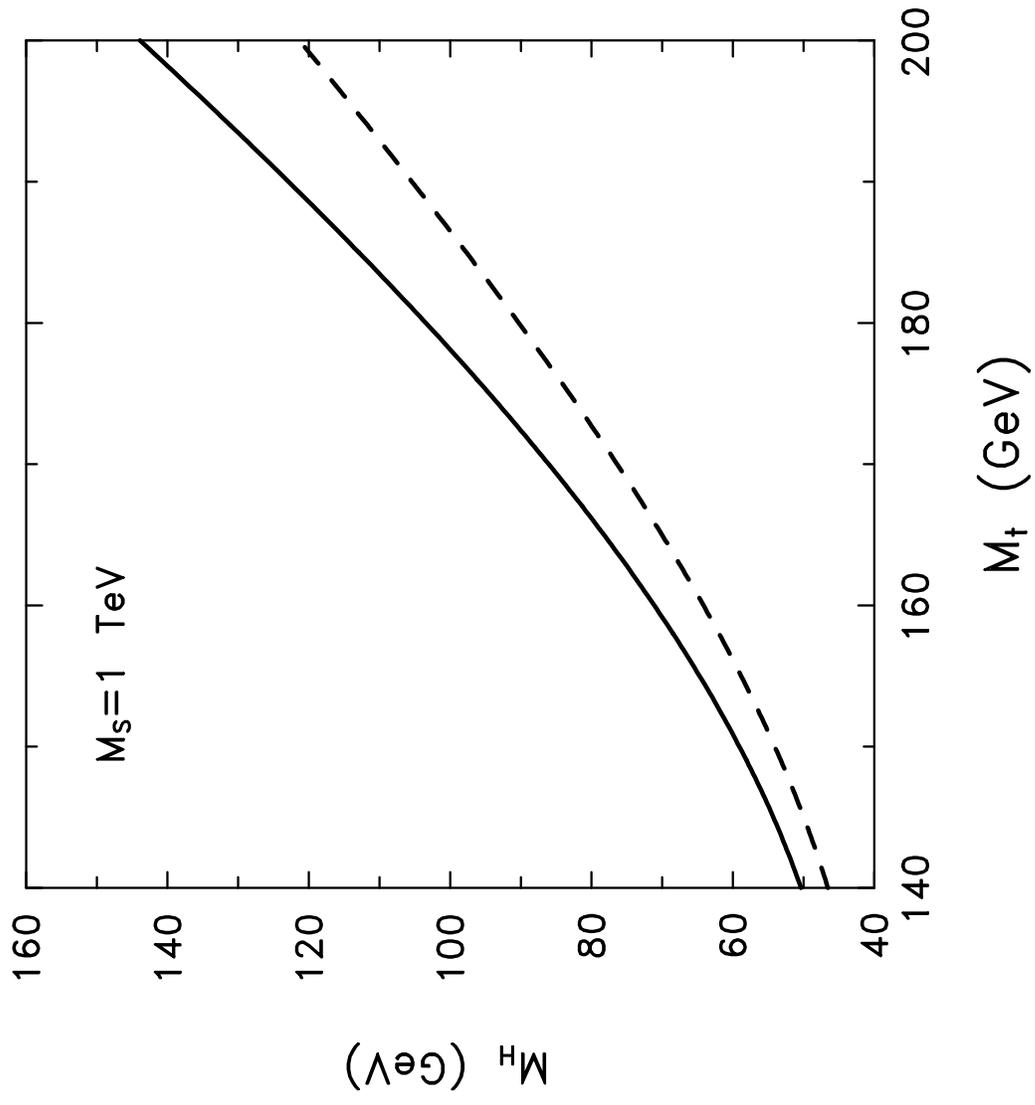
\item[Fig. 9] Plot of $M_H$ as a function of $M_t$ for $M_S=1$ TeV
and $\tan\beta$ determined from the fixed point condition
(\ref{mtfp}). The solid line corresponds to maximal mixing,
$X_t^2=6M_S^2$ and the dashed line to $X_t=0$.
\end{description}

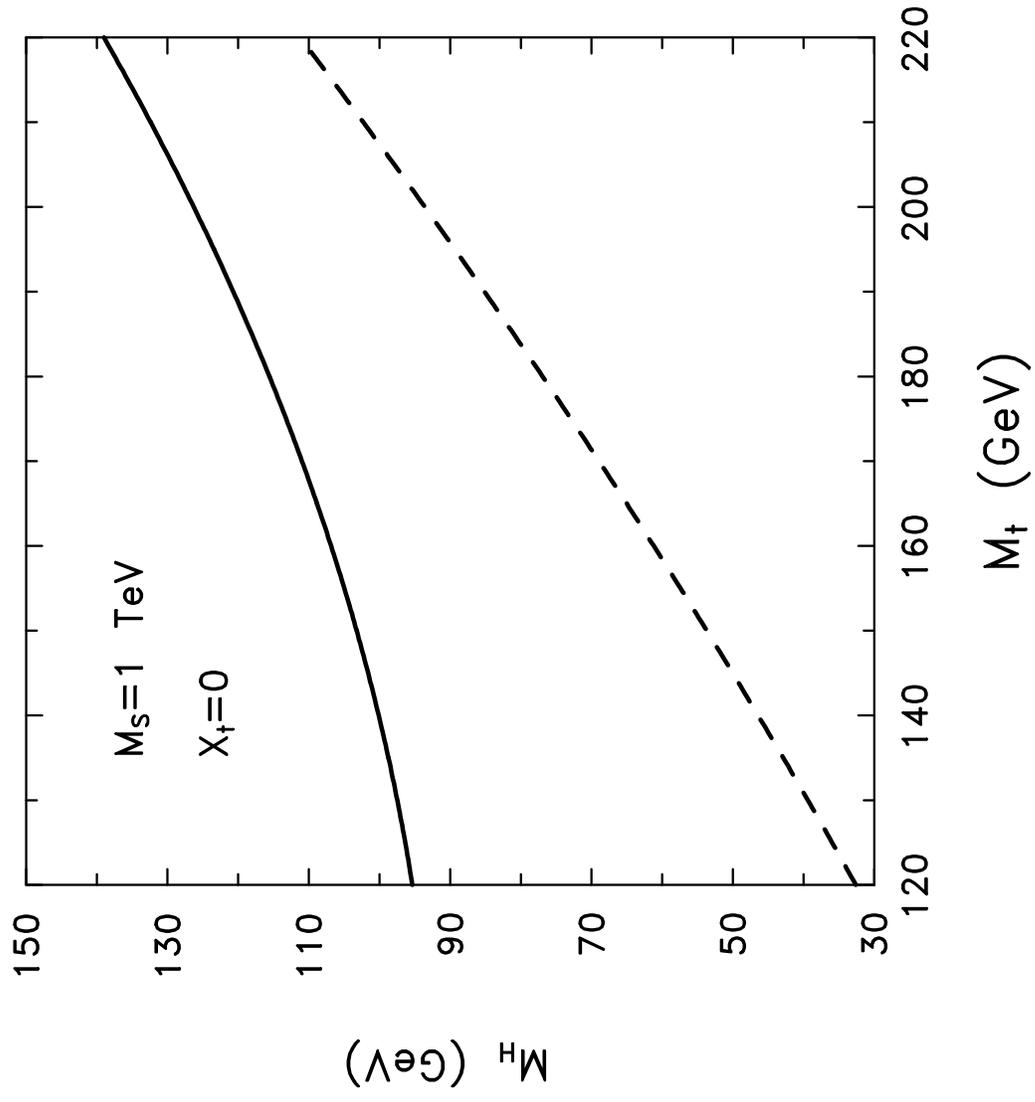
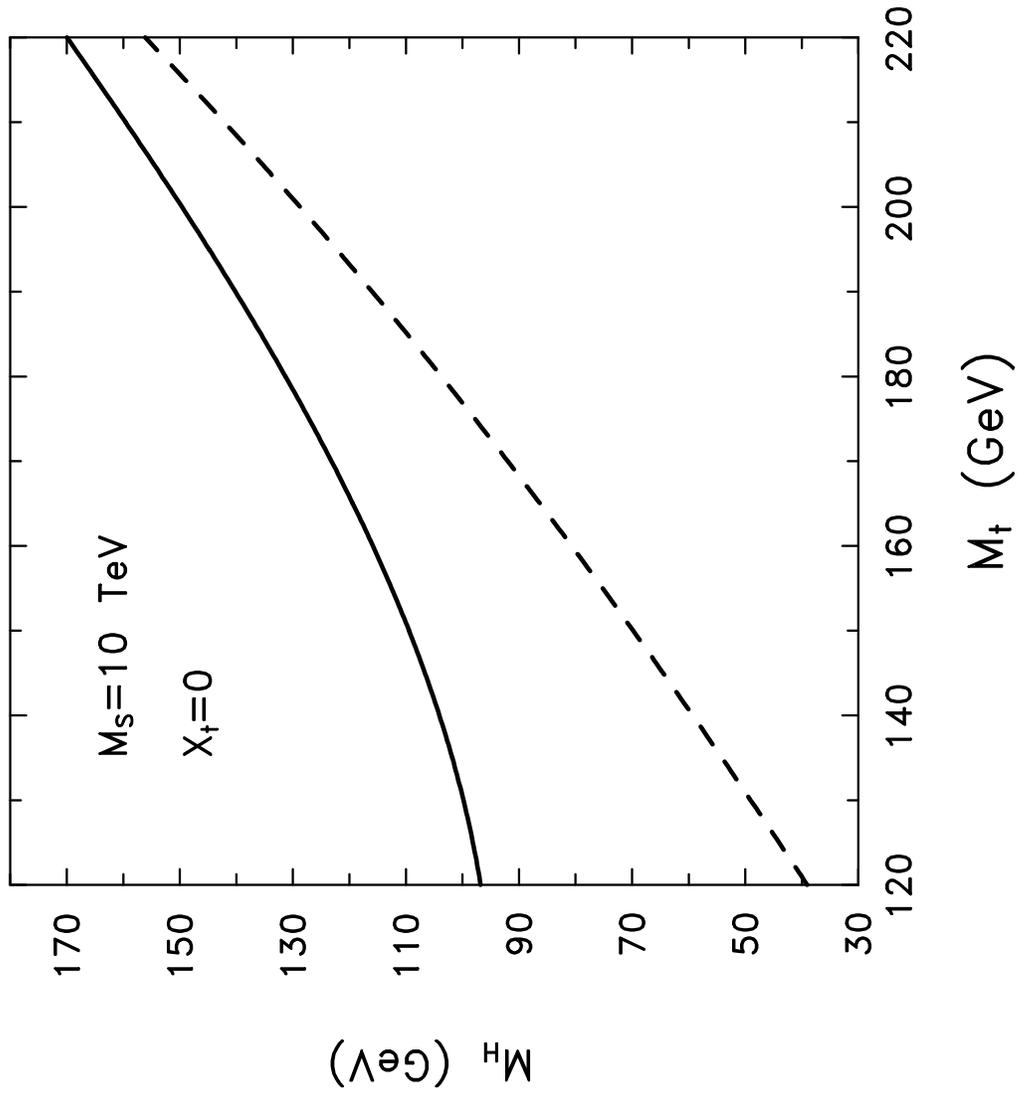
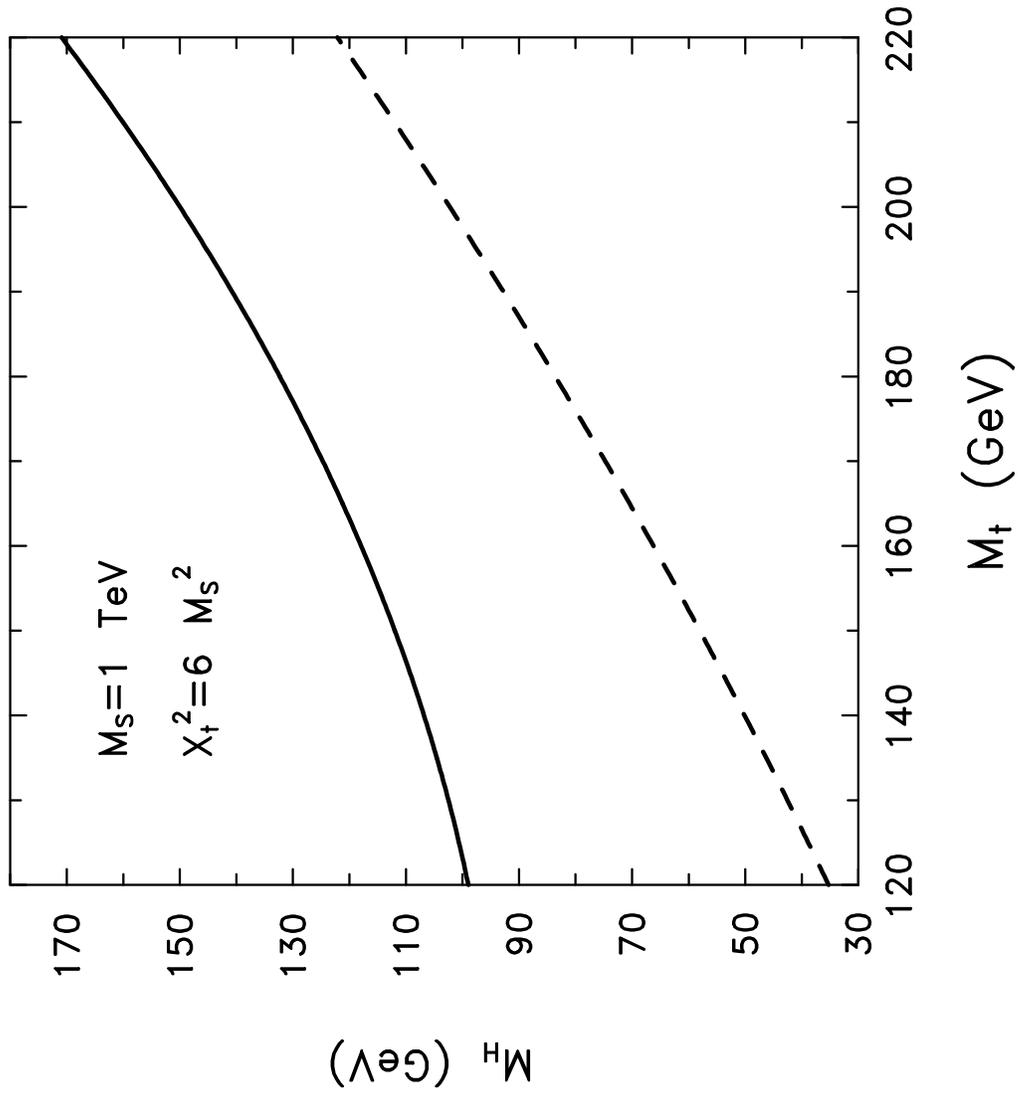
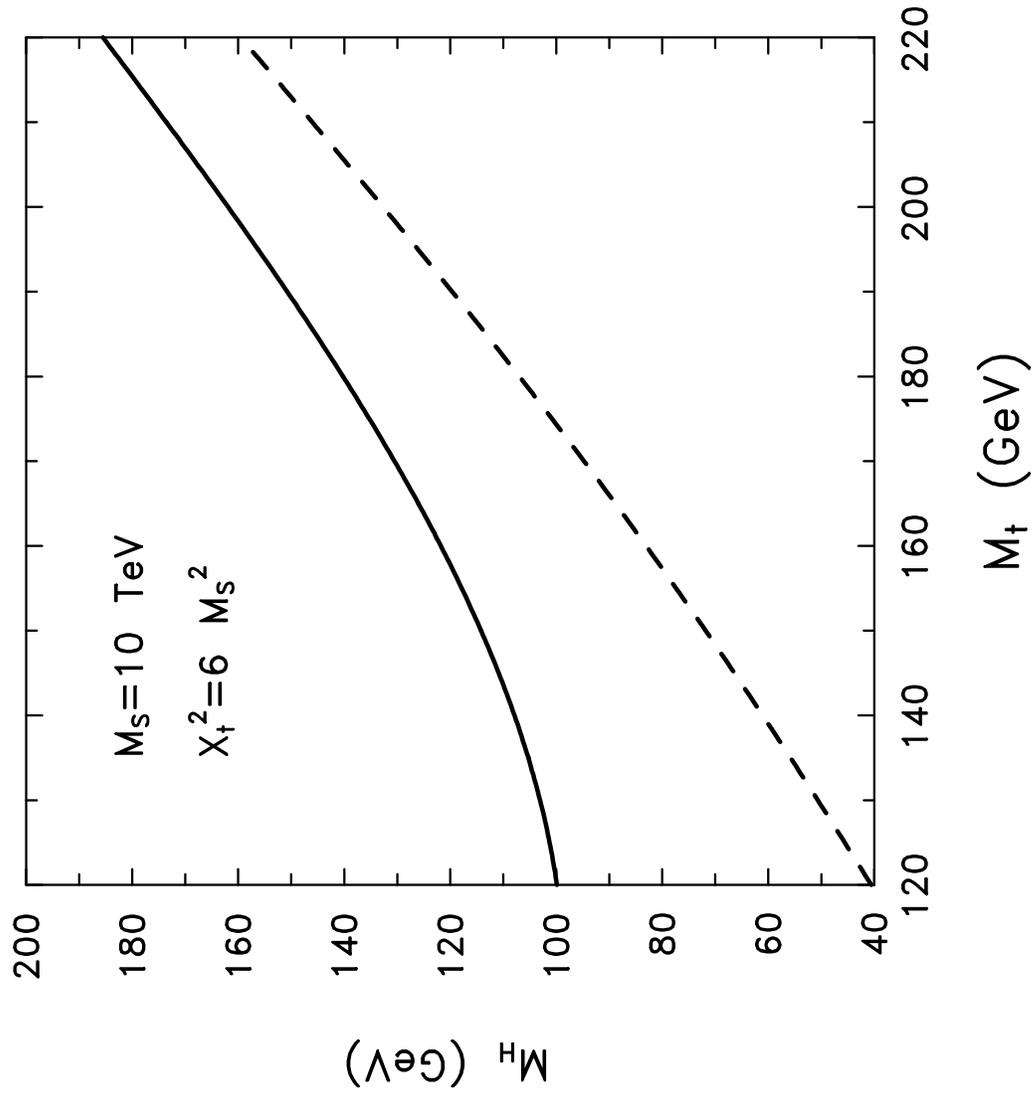
\end{document}